\documentclass[preprint2]{aastex}
\usepackage{graphicx,lineno}
\usepackage{color,float}
\usepackage{xspace}
\usepackage{amssymb}
\usepackage{amsmath}
\usepackage{multicol,multirow,booktabs}
\usepackage{natbib}
\def\msun{\rm{\,M_{\odot}}}

\def\msunh{\rm{\,M_{\odot}{\it h}^{-1}}}

\def\ergs{\rm{\,erg~s^{-1}}}

\def\sxu{\rm{\,cm^2}{gr^{-1}}}
\def\dms{\rm{\,gr{\,cm^{-2}}}}

\def\kms{\rm{\,km~s^{-1}}}

\def\ctsn{\rm{\,counts~arcsec^{-2} }}
\def\pc{\rm{\,pc}}
\def\kpc{\rm{\,kpc}}
\def\mpc{\rm{\,Mpc}}

\def\XDM{\rm{X-DM}}
\def\BDM{\rm{BCG-DM}}

\def\concI{\rm{c^{NW}_{200}}}
\def\concII{\rm{c^{SE}_{200}}}
 \def\dxse{d^{SE}_{X-DM}}
 \def\dmcl{d^{obs}_{DM}}
 \def\tpcl{t^{obs}_{p}}

\newcommand{\gr}{\kern 2pt\hbox{}^\circ{\kern -2pt K}} %  ====> GRADI KELVIN

\def\Log10{{\rm~Log_{10}}}

\def\lcdm{$\Lambda$CDM}
\def\om{\Omega_m}
\def\oml{\Omega_{\Lambda}}

\def\glens{$g_T(\theta)$}
\def\glJ{$g^J_T(\theta)$}
\def\glK{$g^K_T(\theta)$}
\def\Xmph{$X_{mph}^{obs}$}
\def\mnw { M^{NW}_{200}}
\def\mse { M^{SE}_{200}}

\newcommand{\ltsima}{$\; \buildrel < \over \sim \;$}
\newcommand{\simlt}{\lower.5ex\hbox{\ltsima}}
\newcommand{\gtsima}{$\; \buildrel > \over \sim \;$}
\newcommand{\simgt}{\lower.5ex\hbox{\gtsima}}
\newcommand{\be}{\begin {equation}}
\newcommand{\ee}{\end {equation}}

\def\aap{A\&A}
\def\apj{ApJ}

\def\apjl{ApJ}
\def\mnras{MNRAS}

\def\aj{AJ}

\def\physrep{Phys. Rep.}

\def\apjs{ApJS}

\def\prd{Phys. Rev. D}

\defcitealias{Valda24}{V24}

\shortauthors{Valdarnini R.}
\begin{document}

%%%%%%%%%%%%%%%%%%%%%%%%%%%%%%%%%%%%%%%%%%%%%%%

\topmargin -1.3cm
% NOTE THAT THIS WILL PLACE THE LAYOUT CORRECTLY ON ASTRO-PH
% PLEASE LEAVE IN

%%%%%%%%%%%%%%%%%%%%%%%%%%%%%%%%%%%%%%%%%%%%%%%%
%\shorttitle{New variable sources in LMC}

%\shortauthors{R. Valdarnini}

%--------------------------------------------------

\title{ Hydrodynamical simulations  of the merging
cluster El Gordo  in a two-state self-interacting dark matter 
scenario}

\author{R. Valdarnini$^{1}$ }
\affil{$^1$SISSA, Via Bonomea 265, I-34136, Trieste, Italy}
\email{valda@sissa.it}

%------------------ ABSTRACT ----------------------
\begin{abstract}
 A large suite of N-body/hydrodynamical simulations  is used to demonstrate 
that the physical properties of the  merging  cluster  El Gordo are well 
reproduced  within a two-state self-interacting dark matter scenario.
Our findings show that, in addition to elastic scattering,  the presence of an 
inelastic, endothermic, up-scattering channel is essential to satisfy 
observational constraints derived from the measured weak lensing profiles of
the two colliding clusters.
We  find that the observational properties of El Gordo are
adequately matched by  our self-interacting dark matter model  within 
 the following interval of parameters:
 an  elastic cross-section in the  range of $ \sim 4 \textendash 6 \sxu$,
and an  up-scattering channel bounded by an inelastic cross-section
of approximately $ \sim 2 \textendash 4 \sxu$ with  a threshold velocity of 
$\sim 1200 \textendash 1600 \kms$, respectively.

We argue that our conclusions depend critically on current uncertainties  in
the primary mass   and the projected separation  $d_{DM}$.
In fact, our findings suggest that if  error estimates  in weak 
lensing measurements were  significantly reduced - with 
future observations showing evidence of  a low primary mass 
($\simlt10^{15} \msun$) and a larger projected separation 
$ d_{DM} \simgt 740 \kpc$ - it would  then be difficult for the 
proposed self-interacting  scenario  to be reconciled  with observational 
constraints.
We further suggest  that  future lensing surveys of massive, high-velocity 
merging clusters could be   decisive in  testing the  proposed dark matter 
scenario.
\end{abstract}
%--------------------------------------------------

%--------------------------------------------------

%
%\cleardoublepage

\section{Introduction}
\label{sec:intro}

According to the standard cosmological paradigm, the dynamics of structure
formation in the Universe is driven by gravity and proceeds hierarchically
from the bottom up.

In this framework most of the matter content in the Universe ($\sim 80 \%$) 
is made up of an unknown dark matter (DM), which hypothetically consists of 
collisionless cold particles (CDM) that interact only via gravity.
This CDM cosmological model, with the inclusion of a cosmological constant
 to achieve a  spatially flat Universe (\lcdm), has been shown to  successfully
 describe structure formation over a wide range of scales.
 Nonetheless, the model suffers from a number of problems on small scales 
 \citep{Wei15,Tulin18}. These issues, such as  the core-cusp problem, the 
 missing satellites problem, and the  too-big to fail problem,  have led many 
 authors to consider viable alternatives to the standard CDM model 
 \citep{Wei15}.

 The possibility that DM is self-interacting (SIDM) is 
 one of the proposed  solutions to solve  the problems of CDM at small scales 
\citep[see][for a review ]{Tulin18}.  In this scenario, the energy and momentum of 
DM particles will be 
redistributed in the presence of  collisions.  Thus, self-interactions of
DM particles  in the 
inner regions of DM halos will reduce central DM densities  and alleviate 
the cusp-core problem.

Other astrophysical systems in which  DM collisionality is expected to exhibit 
significant signatures are collisions between massive galaxy clusters.
Major mergers between  galaxy clusters 
\citep[see][for a review ]{Molnar16} are very  energetic events, 
with collisional energies  as high as  $ \sim 10^{63} - 10^{64} {\rm{\,erg}} $.
They can then be used as natural particle colliders 
to probe DM  collisional properties.

In a standard (collisionless) CDM scenario an important observational 
signature in   merging clusters is the spatial separation, after the 
first pericenter passage, between the 
collisionless (DM and galaxies) components from the X-ray emitting gas.
If DM is self-interacting, an additional offset is expected  
 to be present between the DM itself and the galaxy component.
Measurements of spatial offsets in merging clusters have then been used by
various authors  \citep{Tulin18} to place constraints on SIDM models.

 N-body/hydrodynamical simulations  are a natural tool to correctly model the 
 non-linearity of the physical phenomena taking place during the merging 
 process. Specifically, idealized binary cluster mergers are used 
 to follow the collision evolution of two individual halos 
 initially separated and at equilibrium. This approach 
 allows us to study a single merger event under controlled initial conditions
 \citep{Molnar16}.
 A well-known example of a merging cluster is the Bullet Cluster 
 \citep{Spr07,Mas08,La14}.  Other best studied examples include the
 `El Gordo' cluster \citep[ACT-CL J0102-4915,][] {Donnert14,Molnar15,Zh15} and 
  the Sausage Cluster \citep{Donnert17,Mol17}.

 In particular, the El Gordo cluster at $z=0.870$ is an example of a  rare 
 merging system colliding at a high speed ($2,000 \kms  \simlt V 
 \simlt 2,500 \kms$).
  In accord with dynamical mass estimates \citep{Men12}, 
 weak lensing \citep[WL;][]{Jee14} and strong lensing (SL) studies 
 \citep{Zi13} estimate a total cluster mass  of about  
 $\sim 2 \cdot 10^{15} \msun$.
 As has been pointed out by several authors \citep{Men12,Jee14,As21,As23}, 
 the existence at high redshifts of such a  high-speed major merging cluster
 is a serious challenge to the standard \lcdm~cosmological model.

 Multifrequency observations reveal the existence of 
 two subclusters \citep{Jee14}, which are commonly referred to as the 
northwestern (NW) and southeastern (SE) components 
\citep[see, e.g., Figure 1 of ][]{Ng15}. 
The  projected separation  of  these two subclusters is 
 $ d \sim 700 \kpc$, with an estimated  mass ratio 
  $\sim 2:1$ \citep{Men12}.

 X-ray data analysis shows that the merging cluster possesses  a distinct 
 X-ray morphology, suggesting  a  merging  plane  that is  very close to 
 the plane of the sky  \citep{Men12}.
 The total X-ray luminosity  is estimated to be  
  $L_X\sim 2\cdot 10^{45} \ergs$ in the $0.5-2$ keV band \citep{Men12}, 
  with the majority of it coming from an emission peak localized in the SE
  region.  A noteworthy feature of the X-ray emission is the presence of 
  a distinctive, twin-tailed, cometary morphology that extends beyond 
  the SE X-ray emission peak.

  Moreover, based on multifrequency observations, 
the peak locations   of the various  mass components 
  can be considered among the most significant observational features of 
  the El Gordo cluster \citep{Men12,Jee14, Zi13,Die20,Kim21}.

As expected in cluster collisions, the X-ray peak of the SE cluster is 
spatially separated  from  the corresponding DM peak. 
However, its spatial location is more distant from the system's center-of-mass 
than the corresponding DM mass centroid. 
This is at variance with what is predicted from dissipative 
arguments and what is seen, for example, in the Bullet Cluster. 
It is worth noting that WL studies report this offset between the X-ray peak 
and the DM centroid at $2 \sigma$ \citep{Kim21}.
Moreover,  the brightest cluster galaxy (BCG),
which is part of the SE cluster \citep{Men12},   is  also spatially offset 
from the SE mass centroid.  It should be stressed that the presence of 
DM self-interactions in major cluster mergers  \citep{Kim17} is expected to 
lead to  galaxy-DM offsets.

A scenario consistent  with these  observational findings 
indicates that El Gordo consists of two massive, high redshift clusters,  
colliding at high speed ($ \simgt 2,000 \kms$) in the plane of the sky.
The preferred model is an outgoing scenario \citep{Men12,Jee14}, 
where the two clusters  have passed the first core passage and are now moving 
away from each other.

In this scenario, the motion of the  SE cool gas core  
through the intracluster medium (ICM) of the NW cluster 
generates the wake and the two-tailed  X-ray morphology,   
as observed in the X-ray maps.  Regarding the NW cluster,  an X-ray emission peak
is absent because it was destroyed during the 
collision with the compact SE gas  cool core.  

N-body/hydrodynamical simulations  have been used 
 \citep{Donnert14,Molnar15,Zh15,Zh18,Valda24}  to constrain models of the 
 El Gordo merger,  aiming to match the set of the different 
 multifrequency  observations.  In particular, recent simulations 
 \citep[][hereafter \citetalias{Valda24}]{Valda24}
 showed that, in a collisionless 
 CDM scenario, the observed X-ray morphology is well reproduced 
by  an off-axis merger with a primary mass 
between $\sim 10^{15} \msun$ and $\sim 1.6 \cdot 10^{15} \msun$,  
initial  collision velocities $2,000 \kms  \simlt V \simlt 2,500 \kms$  and 
impact parameters $600 \kpc \simlt P \simlt 800 \kpc$. 

However, the main shortcoming of a standard CDM merger 
is the difficulty in explaining the observed spatial separations
between the different mass components \citep{V25}.
Specifically, for the SE cluster, the DM centroid is observed trailing 
the X-ray emission peak, in addition to the presence of a BCG to DM offset.
On the contrary, these observational features  are naturally reproduced by 
the merger process if DM is self-interacting, given that the possibility of 
 galaxy-DM offsets is a specific prediction of SIDM in major mergers.

 Simulations of merging clusters in an SIDM framework 
 have been studied by  many authors \citep{Mark04,Kah14,Kim17,Rob17,
 Rob19,ZuH19,Harvey19,Fis22,Fis23,Cross24,Sab24,Sirks24}.
In particular, for the merging cluster El Gordo,
we  recently presented results from SIDM merger simulations 
with velocity-independent elastic cross-sections \citepalias{Valda24}.

The simulations showed that an SIDM merger model with an 
elastic cross-section of approximately  $\sigma_{DM}/{m_X}  \sim 4  \sxu$ 
is able to match the observed spatial separations between the 
 different centroids.
Moreover, the mean relative  radial velocity between the
two BCGs  is   on the order of several
hundred $\kms$, in accord with the observational range.

Despite these interesting features, the proposed SIDM merger model for the 
El Gordo cluster is not entirely free of problems \citep{V25}. 
The most serious tension is the inconsistency between the reduced tangential 
shear lensing profiles of the halos, as predicted by the simulations, and the 
measured profiles.  The shear lensing profiles extracted from the simulations
consistently tend to zero at small angles \citep{V25}, in accord with the 
cored density profiles of the post-collision DM halos 
as predicted by  SIDM.
This is in contrast with the observed lensing profiles, 
which  exhibit a diverging behavior at small angles, as 
predicted by an NFW halo model.

 Motivated by these difficulties, we present here results from a 
 suite of 
  N-body/hydrodynamical SIDM simulations aimed at studying the merging cluster 
  El Gordo.  These simulations are based on SIDM models more general  than
   the model  previously employed \citepalias{Valda24}, which assumed  
   elastic scatterings between two SIDM particles.

   Specifically, a natural extension to an elastic SIDM  model is to assume
   the possibility of inelastic scattering between DM particles.
   This possibility is theoretically motivated in a two-state DM model
   \citep{Sch15}, where DM particles can scatter inelastically between 
   a ground state ($\chi_1$) and an excited state ($\chi_2$).
    In such a model, kinetic energy  is not conserved during the 
    collision between the two DM particles. An endothermic process occurs 
    when there is an up-scattering   $\chi_1 \rightarrow \chi_2$. Conversely, 
 the  process is exothermic in the case of a 
    down-scattering $\chi_2 \rightarrow \chi_1$.

Thus far, much of the attention for a two-state inelastic DM model 
  has been dedicated to the study of galactic- or dwarf-sized halos
   \citep{Todo19,Vog19,Alv20,Huo20,Chua21,Todo22,ON23,Leo24,Shen24,Kim25}.
   This is because a two-component inelastic SIDM 
   model can mitigate the well known  small-scale problems, such as core 
   formation, when down scattering events are present
\citep[see, e.g.,][]{Chua21,ON23}.

In contrast, the impact of inelastic SIDM on cluster scales 
has received lesser attention \citep{Todo22,Huo20}.
In this work, we assume that our simulations include both  elastic and 
inelastic scattering of DM particles.
Specifically, in addition to elastic scattering, we will consider the 
possibility of up-scattering events occurring between two DM particles.

Our interest in such a two-state self-interacting DM  model 
is strongly motivated by the presence of DM halos with a  cored density profile
in our SIDM merger model for the El Gordo cluster \citepalias{Valda24}.
At small angles  the corresponding shear lensing profiles 
are consistently found to tend to zero, in sharp contrast with the measured 
profiles. These are found to diverge at small angles, in accord with DM halos 
that have  a cuspy radial density behavior as $r \rightarrow 0$.

We suggest here that this inconsistency could be solved by adopting the 
proposed two-state self-interacting DM  model 
as a possible  SIDM merger model for the El Gordo cluster.
The primary physical motivation for this generalization 
is that,  in an endothermic DM self-interaction scenario,
 up-scattering  events are expected to lead to 
an energy loss of the ground state DM particles, which in turn  result
in steep inner density profiles \citep{Huo20,ON23}.

The opposite effect happens in the presence of exothermic reactions. In such a 
case down-scattering events will boost the velocities of ground state particles, 
thus helping to create cored density profiles in DM halos
   \citep{Todo19,Vog19,Chua21,Todo22,ON23,Leo24,Kim25}. In this work we will 
   restrict our investigation to the study of  endothermic models 
   in an inelastic DM self-interaction scenario.

 The paper is organized as follows: Section \ref{sec:sims} provides a brief description
 of the 
  simulation setup:
  the construction of the merging initial conditions is outlined in 
  Section \ref{subsec:icsetp}, while the numerical implementation of  
  DM self-interactions  is presented in Section \ref{subsec:icsidm}. 
  Section \ref{subsec:imag}  describes the procedures used to 
  construct mock X-ray maps, and   Section \ref{subsec:lensprof}
   introduces the methodology  for extracting  tangential shear 
  lensing profiles from the simulations.
 In Section \ref{subsec:obsvcon}  we discuss a number of observational
 properties of the El Gordo clusters used to  constrain  
  the initial parameters of our SIDM merger simulations.
   Finally, Section \ref{subsec:mrgmodels}  discusses 
 the choice of the optimal initial condition setup for some 
 of these parameters.

The results are presented in Section \ref{sec:results}:    
 Section \ref{sec:opt} first  discusses  the dependency 
 of key observational properties on various simulation parameters, while
  Section \ref{sec:sidm}  presents the results from those fiducial merger
 simulations  that, within a two-state endothermic  SIDM scenario, are found to
 best reproduce the main observational features of El Gordo.
Finally, Section \ref{sec:discuss} summarizes our main conclusions.
Throughout this work  we use a concordance \lcdm~ cosmology,  with
$\om=0.3$, $\oml=0.7,$ and Hubble constant 
$H_0=70\equiv 100h$\,km\,s$^{-1}$\,Mpc$^{-1}$.

%%%%%%%%%%%%%%%%%%%%%%%%%%%%%%
%\clearpage
\begin{table*}
\caption{ Initial collision parameters of the  off-axis merger 
simulations of Sections \ref{sec:results}. $^{a}$
}
\label{clparam.tab}% 
\centering
\begin{tabular}{ccccccccccc}
\hline
ID$_{coll}$ & $M^{(NW)}_{200}$ $[\msun]$ & $M^{(SE)}_{200}$ $[\msun]$ & 
	$r^{NW}_{200}~[\mpc]$ & $c^{NW}_{200} $ & $ r^{SE}_{200}~[\mpc] $ 
	& $V~ [\kms] $ & $P~[\kpc] $ & $d_{ini}~[\mpc]$ \\
\hline
Bf & $1.6 \times 10^{15}$ & $ 6.9 \times 10^{14}$ & 1.74 & 2.5 & 1.32 & 2500 & 600 & 6.12 \\
Bl & $ 1 \times 10^{15}$ & $ 6.5 \times 10^{14}$ & 1.49 & 2.6 & 1.29  & 2000 & 600  & 5.56 \\
Bn & $ 1 \times 10^{15}$ & $ 6.5 \times 10^{14}$ & 1.49 & 2.6 & 1.29  & 2500 & 600  & 5.56\\
Bw & $ 8.5 \times 10^{14}$ & $ 5.2 \times 10^{14}$ & 1.41 & 2.64 & 1.2 & 2000 & 600  & 5.22\\
By & $ 8.5 \times 10^{14}$ & $ 5.2 \times 10^{14}$ & 1.41 & 2.64 & 1.2 & 1800 & 600  & 5.22 \\
%Br & $ 1 \times 10^{15}$ & $ 6.5 \times 10^{14}$ & 1.49 & 2.6 & 1.29  & 3000 & 600  \\
%Bq & $ 1 \times 10^{15}$ & $ 6.5 \times 10^{14}$ & 1.49 & 2.6 & 1.29  & 2500 & 300  \\
%Bm & $ 7.5 \times 10^{14}$ & $ 5.2 \times 10^{14}$ & 1.36 & 2.66 & 1.2 & 2500 & 600  \\
\end{tabular}
\begin{flushleft}
{\it Notes.} {$^a$ Columns from left to right:
model label  of the collision parameters, halo mass
$M^{(NW)}_{200}$ of the primary, 
halo mass $M^{(SE)}_{200}$ of the secondary , 
cluster radius  $r^{NW}_{200}$ of the primary,
 halo concentration parameter $\concI$ of the primary as given by
Equation (\ref{cfit.eq}), cluster radius  $r^{SE}_{200}$ of the secondary, 
initial collision velocity $V$, collision impact parameter $P$,
 separation $d_{ini}= 2(r^1_{200}+r^2_{200})$ between the cluster centers
when the  collision parameters $\{V~,P~\}$ assume the listed values.
}
\end{flushleft}
\end{table*}
%   Bf Bg Bh  conc=2.5 r_200=1.74 r_s=0.698    II conc=2.682  r_200=1.32
%  Bk Bl conc=2.6 r_s = 0.577
%%%%%%%%%%%%%%%%%%%%%%%%%%%%%%%%%%%%%%%%%%%%%%%%%%%%%%%%%%%%%%%%%%

\section{Method}
\label{sec:sims}
As in \citetalias{Valda24},  our idealized binary cluster merger simulations 
were carried out employing a parallel N-body/SPH treecode. The SPH code adopts
an entropy-conserving formulation,  in which gradient errors are strongly 
reduced by using an integral approach. We refer to  \citet{V16} for a thorough 
study of the hydrodynamical performances of the code.
We now briefly describe the initial condition setup  of the merger simulations;
see  \citetalias{Valda24} for a more extensive description of  the adopted 
procedures.

\subsection{Initial conditions }
\label{subsec:icsetp}
The two merging clusters consist of a primary with mass  $M_1$ 
and a secondary with mass $M_2$,  we denote the mass ratio as  $q=M_1/M_2 \geq 1 $.    
 The mass of a  cluster is defined  as $M_{200}$,  the cluster mass 
 within   the radius $r_{200}$.  This radius corresponds to the radius within which 
 the average cluster density 
 at  $z=0.87$,  the redshift of the El Gordo cluster,
is  $ 200$ times the cosmological critical density $\rho_c(z)$:

 \begin{equation}
M_{200}=\frac{4 \pi}{3} 200  \rho_c(z) r_{200}^3~.
 \label{mcl.eq}
 \end{equation}

Before the merging, the two colliding clusters are represented by 
 two individual halos. The mass components  of each halo consist of DM, gas,  
 and  a stellar component. We assume that the halos are in 
 hydrostatic equilibrium and construct a particle realization of position and 
 velocities  for the different mass components
 according to the corresponding density profiles.

\subsubsection{ Halo density profiles   }
\label{subsec:icdm}
We have chosen an NFW profile to model the  DM density profile of the halos:
\begin{equation}
\begin{array}{llll}
\rho_{DM}(r)&=&\dfrac{\rho_s}{r/r_s(1+r/r_s)^2}\, ,  &   0\leq r\leq r_{200} \, , \\
 \end{array}
 \label{rhodmins.eq}
 \end{equation}
where  $ r_s=r_{200}/c_{200}$ is the scale radius and $c_{200}$ is the concentration 
parameter of the halo.  This is determined  according to 
the $c-M$ relation of \citet{Du08} :

 \begin{equation}
	 c_{200}= 5.71 \left(\frac{1}{1+z}\right)^{0.47} 
	 \left( \frac{M_{200}} {2\cdot10^{12} \msunh}
\right)^{0.084}~.
 \label{cfit.eq}
 \end{equation}
The DM cluster density profile is then specified  once the mass $M_{200}$ and 
 the concentration parameter $c_{200}$ are given.

\begin{table*}
\begin{center}
\caption{ Initial gas density profile parameters  adopted for 
the various merger simulations of Sections \ref{sec:results}. $^{a}$ }
%\begin{tiny}
%\scalebox{0.9}{
\label{subcl.tab}% 
\begin{tabular}{c|cc|cc|c|c}
\toprule
\toprule
\multicolumn{7}{c}{}\\
 \multicolumn{1}{c}{ }&
 \multicolumn{2}{c}{{\rm cluster gas fractions} }&
% \multicolumn{1}{c}{ }&
 \multicolumn{2}{c}{{\rm gas core radii} }&
% \multicolumn{1}{c}{ }&
\multicolumn{1}{c}{{\rm gas profile of  the SE cluster}}  &
 \multicolumn{1}{c}{{\rm halo concentration of \linebreak the SE cluster}} \\ 
%\cmidrule{rl}{2-3} 
\cline{2-3}
\cline{4-5}
\cline{6-6}
\cline{7-7}
 \multicolumn{7}{c}{} \\
%\midrule
ID$_{gas}$  & $f^{NW}_{g}$ & $f^{SE}_{g}$  & $r^{NW}_c[\kpc] $  & 
	$r^{SE}_c [\kpc] $   &   model & $c_{200}^{SE}$  \\   \hline
Br\_a   &   0.1 & 0.1 &   200  & 164  &  Burkert &  2.68 \\   %F1
Br\_b &   0.12 & 0.14 &   290  & 164   &  Burkert &  2.68 \\    
Br\_c    &   0.12 & 0.14 &   170  & 160  &  Burkert &  2.7 \\  %F2  
Br\_d     &   0.16 & 0.14 &   250  & 135  &  Burkert &  3.2 \\ %F3+4   
Br\_e   &   0.16 & 0.14 &   290  & 135    &  Burkert &  3.2 \\  %F6  
Bm\_a   &   0.16 & 0.14 &   250  & 161  &  $\beta$-model &  3.2 \\   %F6
Bm\_b   &   0.16 & 0.16 &   210  & 135 &  $\beta$-model &  3.2 \\    % F8
Bm\_c  &   0.16 & 0.16 &   210  & 162 &  $\beta$-model &  3.2 \\    % F9
\hline
\hline
\end{tabular}
\begin{flushleft}
{\it Notes.} {$^a$ Columns from left to right:
model label for the specified  cluster gas density profiles,  
primary and secondary cluster gas mass fractions $f_g$ at  $r_{200}$,  
 gas core radii for both the clusters, 
adopted model for the gas density profile of the secondary,
 halo concentration parameter $\concII$ of the secondary.
Additional notes:  We adopt $\beta=1$ for all the $\beta$-models;
in all cases, the   gas density profile  of the primary is represented by 
a $\beta$-model.  For the SE cluster,  $\concII=3.2$ is the 
best fit value taken from  Table 2 of \citet{Kim21}.
}
\end{flushleft}
%\end{tiny}
\end{center}
\end{table*}

%Unless otherwise specified, this is determined  by the $c-M$ relation (4) 
%of \citet{Du08}, 
%with the coefficients  taken from the second row of  their Table 1.
We use an  exponential cutoff to model the DM density profile for $r>r_{200}$,
setting a  scale length of $r_{decay}=0.2 r_{200}$ and extending the profile 
up to  $r_{max}=2 r_{200}$ \citep{VS21}.

 We construct  a particle realization of the DM density profile 
 by first evaluating the  cumulative DM mass profile,  $M_{DM}(<r)$.
 The radius $r$ of a particle is then found by inverting the equation 
 $x=M_{DM}(<r) /M_{DM}(<r_{max})$, 
 where $x$ is a uniform random number in the interval $[0,1]$.

Finally, a standard acceptance-rejection method is used to obtain
the particle speed $v$ of a particle at the position $r$ \citep{Drakos17}.
This is done  by numerically calculating the DM distribution function 
$f_{DM}(\mathcal{E})$ over a range of energies.
After the radius $r$ and the speed $v$ of the particles are determined, 
the directions of the particle position and velocity vectors are 
chosen isotropically.

We adopt the following functional forms for the gas density profiles of the 
halos.
The  gas density profile of the primary is modeled according to 
 a non-isothermal $\beta$-model \citep{Donnert14}:

\begin{equation}
\rho_{gas}(r)= \rho_{0}  \left(1+\frac{r^2}{r_c^2}\right)^{-\frac{3}{2} \beta}~,
 \label{rhogbeta.eq}
 \end{equation}
where $r_c$ is the gas core radius,  $\rho_0$ the central gas density 
and $\beta$ the slope parameter. For the secondary, we  model the gas density 
  using  either a $\beta$-model or a Burkert profile \citep{Bu95}:
\begin{equation}
\begin{array}{llll}
\rho_{gas}(r)&=&\dfrac{\rho_0}{(1+r/r_c)\left[1+(r/r_c)^2\right]}\, ,  &   0\leq r\leq r_{200} \, . \\
 \end{array}
 \label{rhogins.eq}
 \end{equation}
%we will discuss the motivations behind these choices in Section \ref{subsec:mrgmodels}.

 For a given  cluster  gas mass fraction $f_g$ at $r=r_{200}$,
 we find the central  density $\rho_0$  such that  the specified gas profile
 yields the value of $f_g$.
 Finally, we solve for the gas temperature at radius $r$ by using 
  the equation of hydrostatic equilibrium and assuming  an adiabatic index  
 of $\gamma=5/3$ for the gas.

To mimic the presence of BCGs in the merging simulations, 
a  star matter component is initially incorporated in the halos.
The corresponding density profile is approximated analytically, as 
in  \citet{Mer06}.
We set the  masses of the BCGs according to  \citet{Kr18}; 
for the considered  range of halo masses one obtains  
 $M_{\star,BCG}\sim 2.3 \cdot 10^{12}\msun$ ( see,  for example, 
 Table 7 of \citetalias{Valda24}).
The same procedure adopted for DM particles is used to determine
positions and velocities of the star particles.

Finally, we set the masses of DM and gas particles  as described in 
\citet{VS21}.  In SPH, gas particle  masses  must be kept equal
to avoid numerical instabilities \citep{Price2012}. 
Consequently, the numerical resolution  of the simulation is governed by the
mass of the secondary cluster, which is the less massive component of the
binary system: 
 \begin{equation}
m_d\simeq8 \times 10^8 \msun (M^{(SE)}_{200}/2 \times 10^{14} \msun)~.
    \label{mdark.eq}
 \end{equation}
 The mass of the gas particles is rescaled accordingly  as
  $m_g=f_b m_d/(1-f_b))\simeq 0.16 m_d $, where $f_b\simeq 0.16$ 
  is the cosmological  mass gas fraction.
  As demonstrated by the hydrodynamical cluster merger simulations 
  of \citet{VS21}, these mass assignments produced ICM profiles
 that  are fully numerically converged.
Accordingly, a merging simulation with a 
halo mass of  $M_{200}\sim  10^{15}\msun$ has 
  $N_{DM} \simeq 3.4 \times10^5 $   DM particles.  Similarly, 
 a halo gas mass of $M_{g}\sim  10^{14}\msun$    
is resolved with  $N_g \simeq 1.7 \times10^5 $  gas particles.

For each particle the gravitational softening parameter $\varepsilon_i$ 
is implemented   according to the scaling \citep{VS21}

 \begin{equation}
\varepsilon_i =15.8\cdot ( m_i/6.2\times10^8 \msun)^{1/3}\kpc.
  \label{epsd.eq}
 \end{equation}

 Finally, it is worth noting that for merger simulations incorporating a BCG 
 the number of star particles $N_{s}$ is increased by a factor of $4$
 with respect  the scaling set by Equation (\ref{mdark.eq}).
This choice is introduced to ensure that two-body 
relaxation effects remain negligible for the star particles
throughout the simulations.
With this choice of parameters  one obtains 
$\varepsilon_{\star} \sim  9 \kpc$ for 
$M^{BCG}_{\star} \sim 2 \cdot 10^{12} \msun$ (see Table 7 of 
\citetalias{Valda24}).
The gravitational softening length of the star particles 
is therefore consistent with the lower limit 
$\varepsilon_{\star} \sim r_{200}/\sqrt{N_{s}}$ proposed  by 
\citet{Pow03}.
However, it should be kept in mind that  the collision timescales 
here are  $ \simlt 1$ Gyr, whereas the derivation of that lower limit  requires
 numerical heating to be negligible over a Hubble time.

\subsubsection{Initial merger kinematics}
\label{subsec:ickin}

The merging simulations are performed in a Cartesian system of coordinates,
 $ \{x,y,z\}$, with the center of mass of the two clusters at the origin.
The orbits of the two halos are initialized in the $ \{x,y\}$ plane  at $z=0$.
The initial position and  velocity vectors
$ \{{\bf X}^{in},{\bf V}^{in}\}$  of the two cluster centers are set 
following \citet{Zh15} and \citetalias{Valda24}, with the two centers 
 initially separated  by a  distance $d_{ini}=2(r^1_{200}+r^2_{200})$.
The merger  evolution is then  fully determined by the merging 
dynamical parameters $ \{M_1, ~M_2,~ P, ~V\}$, where $V$ is the initial 
relative  velocity of the halos  and  $P$ is the impact parameter of the 
collision.
%The   halo properties and merging parameters of all the  off-axis merger 
%simulations presented in this work are listed in Table \ref{clparam.tab}. 

However, when compared to  previous runs \citepalias{Valda24},  we found the 
two-state SIDM simulations presented here to be computationally more demanding.
For this reason, we modified the initial condition setup of the halos as 
follows.

We consider at an initial time $t_i$ the two clusters as point masses,  with 
positions and velocities  given by 
$ \{{\bf X}^{in},{\bf V}^{in}\}$. We then numerically solve the Kepler problem 
(see Section 2.2.3 of \citet{VS21}) by integrating the orbits of the two 
clusters until a final time $t_f$ such that the separation distance becomes 
 $d_f=d_{ini}/2= r^1_{200}+r^2_{200}$.
 The resulting solution vectors $ \{{\bf X}_{f},{\bf V}_{f}\}$, are then adopted 
 as  the new initial condition vectors 
 $ \{{\bf X}^{in},{\bf V}^{in}\}$  for the merger simulations.
 We found that this procedure has a negligible impact on the 
 final properties of the merging systems, while reducing the computational cost of a 
 simulation by up to a factor of $\sim 2 $. 

 Finally, we assume that the initial NFW profiles of the DM halos 
 in all of our merger simulations match those expected from collisionless DM.
 This assumption is justified, for our considered range of DM 
 cross-sections, because the characteristic timescale for  core 
 formation in an isolated halo ($\sim 1$ Gyr; see, for example,
 Figure A1 of \citetalias{Valda24})
 is significantly  longer than the short interval of time 
 ($\sim 0.15 $ Gyr) required for the two halos to reach the pericenter  and
 collide.

\subsection{Numerical implementation of inelastic SIDM}
\label{subsec:icsidm}

We  consider a two-state SIDM model with two possible states: 
   a ground state $\chi_1$ and an excited state $\chi_2$, 
 with $m_{\chi_2}> m_{\chi_1}$ and a small mass splitting 
 $\delta= (m_{\chi_2}- m_{\chi_1})c^2 < < m_{\chi_1} c^2$ \citep{Sch15,Alv20}.
 For the considered SIDM model , the possible interactions are

\be
\left\{
\begin{array}{llll}
\chi_1+\chi_1 & \rightarrow & \chi_1+\chi_1   &   elastic   \\
\chi_1+\chi_1 & \rightarrow & \chi_2+\chi_2   &   inelastic   \\
\end{array}
\right.
\label{chscatter.eq}
\ee

 An  endothermic inelastic collision
 $\chi_1+\chi_1  \rightarrow  \chi_2+\chi_2$  absorbs energy and 
 can only happen if the relative velocity $v$ between  the two DM particles
  satisfies the following condition 

 \begin{equation}
 \frac{1}{4} m_{\chi_1} v^2 >2 \delta ~.
  \label{thrva.eq}
 \end{equation}

 This condition on the relative velocity can be  reformulated as follows:

 \begin{equation}
v >2 \sqrt{2 \delta/m_{\chi_1}} \equiv 2 V_{th} ~,
  \label{thrvb.eq}
 \end{equation}

 where we have introduced the threshold velocity $V_{th}$.

To numerically implement inelastic DM self-interactions 
in our merging runs,  we follow the probabilistic approach 
of \citet{Vog19}. This numerical scheme can be considered as a 
generalization of the method previously  introduced \citep{Vog12} 
to model elastic scattering between DM particles.

We consider  a generic multistate DM framework where 
each  DM particle $i$ is representative of a  specific state $\alpha$, 
without any contribution from other states. Two DM particles, $i$ and $j$, 
can scatter with one another and modify their states from $\alpha$ and 
$\beta$ to $\gamma$ and $\delta$:

\begin{equation}
\chi_{i}^{\alpha} + \chi_j^{\beta} \rightarrow
	\chi_{i}^{\gamma} + \chi_j^{\delta}~.
 \label{chidm.eq}
 \end{equation}

The masses of the DM particles are $m_i^{\alpha}$ and $m_j^{\beta}$ 
in the states $\alpha$ and $\beta$, respectively, and 
 $m_i^{\gamma}$ and $m_j^{\delta}$  in the states $\gamma$ and $\delta$, respectively.
 In the center-of-mass frame the post-scattering velocities are

\begin{equation}
\left\{
\begin{array}{lll}
	{\bf u_i} &=&\frac{m^{\alpha}_i+m^{\beta}_j}{m^{\gamma}_i+m^{\delta}_j} 
	{\bf V}  + \frac{m^{\delta}_j}{m^{\gamma}_i+m^{\delta}_j}
	{A}_{ij}^{\alpha \beta \rightarrow \gamma \delta } v_{ij} {\bf e}   \\
	{\bf u_j} &=&\frac{m^{\alpha}_i+m^{\beta}_j}{m^{\gamma}_i+m^{\delta}_j} 
	{\bf V}  - \frac{m^{\gamma}_i}{m^{\gamma}_i+m^{\delta}_j}
	{A}_{ij}^{\alpha \beta \rightarrow \gamma \delta } v_{ij} {\bf e}~,  
 \end{array}
\right .
 \label{vdmscatt.eq}
 \end{equation}

 where $v_{ij}=|\bf {v_i}-\bf {v_j}| $ is the relative velocity between
 particles $i$ and $j$, ${\bf V}$ is the center-of-mass velocity of the two 
 particles,
$A_{ij}^{\alpha \beta \rightarrow \gamma \delta } $
 is a dimensionless velocity scale factor, and $\bf e $ is a 
 randomly oriented vector on the unit sphere. The coefficient $ 0 \leq 
A_{ij}^{\alpha \beta \rightarrow \gamma \delta } \leq 1 $
 depends on the scattering channel under consideration, being equal to unity  in the 
 case of elastic scattering and 
$ A_{ij}^{\alpha \beta \rightarrow \gamma \delta } <  1 $
 for endothermic reactions.
 In the latter case it is easily seen that  \citep{Vog19}

 \be
A_{ij}^{\alpha \beta \rightarrow \gamma \delta } = 
\sqrt{ \frac{\mu^{\alpha \beta}_{ij}}{\mu^{\gamma \delta}_{ij}} 
\left(1 + \frac {2 \Delta E^{\alpha \beta \rightarrow \gamma \delta}} 
{\mu^{\alpha \beta }_{ij} v^2_{ij}}\right)}~,
%{\Delta E^{\alpha \beta \rightarrow \gamma \delta} {\mu_{ij}})}
 \label{vtilde.eq}
 \ee

 where we define 
 $ \mu_{ij}^{\alpha \beta}=m_i^{\alpha} m_j^{\beta}/(m_i^{\alpha}+ m_j^{\beta} )$  and
 $ \mu_{ij}^{\gamma \delta}=m_i^{\gamma} m_j^{\delta}/(m_i^{\gamma}+ m_j^{\delta} )$ 
 as  the reduced masses of the two particles.  The masses 
 $ \mu_{ij}^{\alpha \beta}$ and $ \mu_{ij}^{\gamma \delta}$ refer to before 
 and after the scattering event, respectively, and 
 $ \Delta E^{\alpha \beta \rightarrow \gamma \delta}= -2 
 \mu_{ij}^{\gamma \delta} V_{th}^2$ is the energy absorbed during the 
 endothermic scattering reaction.  In accord with our assumption
 $\delta  < < m_{\chi_1} c^2$, we neglect mass differences and assume here 
 $ \mu_{ij}^{\alpha \beta} = \mu_{ij}^{\gamma \delta}$. Moreover, we 
 further assume that all of the DM particles have the same mass.

 In accordance with previous definitions \citep{Vog12,Valda24}, 
 The  pairwise scattering probability of the scattering channel
  $ \alpha \beta \rightarrow \gamma \delta $, 
 within the simulation timestep  $\Delta t_i$, is then defined  as

\begin{equation}
	P_{ij}^{\alpha \beta \rightarrow\gamma \delta} = 
	\delta_{ij}^{\alpha \beta} m^{\beta}_j W_{ij}
	\frac{\sigma_{DM}^{\alpha \beta \rightarrow\gamma \delta } }
	{m_{\chi^{\beta}}} v_{ij} \Delta t_i/2 ~,
 \label{pijmul.eq}
 \end{equation}

 where   $\sigma_{DM}^{\alpha \beta \rightarrow\gamma \delta } $ is the 
 cross-section  of the reaction $\alpha \beta \rightarrow\gamma \delta $, 
  and $W_{ij}= W(r_{ij}, h^{DM}_i) $ is the cubic spline kernel. This is normalized 
  such that  a sphere of radius $2h^{DM}_i$ around particle $i$ contains 
      $N_{nn}=32 \pm 3$ DM neighboring particles.
The factor $2$ in Equation  (\ref{pijmul.eq}) accounts for the other member
of the 
scattering pair. Finally, note that it is possible that not all of the 
$N_{nn}$ neighboring particles around particle $i$  are in the state $\beta$, 
the Kronecker function guarantees that the reaction is effectively 
taking place between particle $i$ in the state $\alpha$ and particle $j$ 
in the state $\beta$.

According to Equation (\ref{pijmul.eq}), 
the total scattering probability $ P_{i}^{s} $
for particle $i$ to interact with any of   its $N_{nn}$ neighbors, through   
 the reaction channel $s: \{\alpha \beta \rightarrow\gamma \delta \}$, is 
 \be
 P_{i}^{s} = \sum_{j} P_{ij}^{s}  
 \label{psumi.eq}
 \ee

  A collision  in the scattering channel 
 $s_i: \{\alpha_i \beta_j \rightarrow\gamma_i \delta_j \}$
  between particle $i$ with one of its 
  neighbors $j$  is assumed to  take place whenever 

 \be
 \sum_{s=0}^{s_i-1} P_i^s< x < \sum_{s=0}^{s_i} P_i^s,
 \label{psum.eq}
 \ee

  where x is a uniform random number in the range $[0,1]$
  and the summation can extend over  all of the scattering channels. Finally,  
  the neighbor $j$ that will be the other member  of the scattering pair is 
  found by solving for 

 \be
 P_i^{s< s_i} + \sum_{k=0}^{k=j-1} P_{ij}^{s_i} < x < 
 P_i^{s< s_i} + \sum_{k=0}^{k=j} P_{ij}^{s_i} ~,
 \label{psumb.eq}
 \ee

 where $ P_i^{s< s_i} = \sum_{s=0}^{s_i-1} P_i^{s} $.

 Hereafter we will use the shorthand notations $\sigma_{11}$
 and $\sigma_{12}$  to denote the scattering cross-sections of the elastic 
 channel $\chi_1+\chi_1  \rightarrow  \chi_1+\chi_1 $ and the inelastic channel 
$\chi_1+\chi_1  \rightarrow  \chi_2+\chi_2$ , respectively.
To summarize, the dependency of the merger dynamical evolution on the chosen 
self-interacting DM model is then fully determined by the  SIDM parameters 
  $\{ \sigma_{11},~\sigma_{12},~V_{th} \}$. 

We will adopt the  notation $\{ \sigma_{11},~0,~0 \}$ for purely elastic SIDM 
simulations   and $\{0\}$  for standard CDM runs.
Moreover, unless otherwise specified, 
the SIDM parameters $\{ \sigma_{11},~\sigma_{12},~V_{th} \}$  
  denote the  DM cross sections in units of $\sxu$ and 
the threshold velocity in units of  $100 \kms$. 
 Note that, by definition, $\sigma_{12}\equiv 0$ for   $v <2 V_{th}$.

As in previous runs \citepalias{Valda24},  the positions and velocities of the 
particles  are integrated according to a standard  block timestep scheme.
 We set  $\Delta t_0=1/200$ Gyr as the largest allowed timestep.
In addition to the standard  timestep limiters, the timestep $\Delta t_i$ of 
a DM particle in the state $\alpha $ must satisfy the following 
additional criterion \citep{Vog19}:

 \be
 \Delta t_i <  \kappa \left [ \rho_i^{\alpha}
 \{ \sigma_{DM}^{\alpha \beta \rightarrow\gamma \delta } (\sigma_i^{\alpha}) /
  m_{\chi^{\beta}} \}_{max} \sigma_i^{\alpha} \right]^{-1}~,
 \label{dtime.eq}
 \ee

 where $\rho_i^{\alpha}$ and $\sigma_i^{\alpha}$  are the local DM density
and velocity dispersion, respectively. 
The notation $\{\sigma_{DM}\}_{max}$ is introduced to 
denote the maximum value of 
$\sigma_{DM}^{\alpha \beta \rightarrow\gamma \delta } $ 
among all the scattering channels. The dimensionless parameter $\kappa$, which 
controls the timestep accuracy, is set to $\kappa=0.01$.
For the simulations presented here, 
 this value ensures sufficiently small  timesteps such that 
 the probabilities (\ref{psumi.eq}) remain  smaller than unity 
 ($P_{i}^{s} <1$), guaranteeing  that the scattering rate is correctly sampled.

 Finally, it is worth mentioning that the parallel implementation of the
 Monte Carlo scheme described here allows an individual DM particle to scatter
 multiple times within a single timestep. This is achieved by performing
 a global sorting of all the scattering pairs at each timestep and 
 consistently updating the particle velocities across all  processors
 after each scattering event. For an in-depth description of the parallel 
 algorithm, see Appendix A of 
 \citetalias{Valda24}.

\subsection{Simulated observations}
\label{subsec:imag}

At any given time during the merging simulations, we extract 
two-dimensional maps of surface mass density, X-ray surface brightness,  
and  Sunyaev-Zel'dovich (SZ) emission. These maps are evaluated 
along the line of sight in the observer's frame  $\{\hat x,\hat y, \hat z \}$, 
where   the cluster merger is taking place in the  plane of the sky
 $\{\hat x,\hat y\}$. 
 The simulation frame $\{ x, y, z \}$  is then related to
 the  observer's frame  by the application of two rotation matrices 
 \citep{Zh15,Valda24}.  Of the two rotation angles, the most important is  
 the projection angle $i$ between the merging axis and the plane of the sky,
 as it determines the observed X-ray morphology.  
 Fiducial merger models were obtained from previous runs \citep{Zh15,Valda24} 
  with the projection angle  set to $i= 30 ^{\circ}$. 
 Here we will extract mock X-ray maps from our SIDM merger models by also 
 considering smaller projection angles: $i=15 ^{\circ}$ and $i= 22^{\circ} $.

The surface mass density  is  calculated by integrating along the line of sight

 \begin{equation}
 \Sigma_m(x,y)= \int_{los}  
\left[\rho_{gas}({\bf x})+\rho_{DM}({\bf x})\right] d z~,
\label{smass.eq}
 \end{equation}
 where $\rho_{gas}({\bf x})$ and $\rho_{DM}({\bf x})$ are the gas and DM
 densities at position $\bf x$, respectively.

 Maps of the X-ray surface brightness  are obtained by integrating 
the X-ray emissivity  $\varepsilon(\rho_g,T_{g},Z, \nu )$    
 along the line of sight and over the energy range $[0.5-2]$ keV
  \citep{Molnar15}: 

 \begin{equation}
% \begin{array}{ccc}
 \begin{split}
         \Sigma_{X}(x,y)  =  &  \frac{1}{4 \pi (1+z)^4}  \int_{los} \,dz  \\
  &        \int \, \varepsilon(\rho_g,T_{g},Z, \nu ) A_{eff}(\nu) \, d \nu  ~,
\end{split}
%\end{array}
\label{sbrx.eq}
 \end{equation}
where  $T_g$ is the gas temperature,  $\nu$ the frequency, $Z$ the 
metal abundance of the gas, and $A_{eff}(\nu)$ the effective area of the 
telescope.  The exposure time of the  mock X-ray maps (\ref{sbrx.eq}) 
 is set to  $t_{exp}=60ks$, and these are then  expressed in counts arc sec$^{-2}$
 \citep{Zh15}.

 We calculate the  SZ surface brightness $\Sigma_{SZ}(x,y)$  at a frequency
of  $\nu=150$  GHz following Equation (5) of  \citet{Zh15}. The resulting
 SZ maps  are smoothed using  a Gaussian kernel with a width of
 $\sigma_{SZ}=270$ kpc, corresponding  to  $\sim0.55^{\prime}$ at
 $z=0.87$.
All of the  2D maps are  evaluated   on a grid of 
 $N_g^2=512^2$ grid points using the SPH smoothing procedure 
 \citepalias{Valda24}.
 Finally, for a given mock image, the spatial location of its centroid 
is found  by applying  a shrinking circle method to the simulation particles.

\subsection{Lensing profiles}
\label{subsec:lensprof}

In the WL regime \citep[see][for a review ]{Um20}  the cluster mass 
distribution at the projected radius $R$ can be probed by the  tangential 
shear $\gamma(R)$. 
This is directly related to the excess surface mass density 
$\Delta \Sigma(R) \equiv \bar {\Sigma}(R) - \Sigma(R)$, where 
$\Sigma(R)$ is the surface mass density,

\begin{equation}
	\Sigma(R)= 2 \int_{0}^{\infty}  \frac{\rho( R^{\prime}) R^{\prime}}
	{ \sqrt{R^{\prime 2}-R^2})}
	dR^{\prime} ~,
 \label{Sigma.eq}
 \end{equation}

  and within the same radius $R$  
 the averaged  surface mass density is accordingly defined as 

 \begin{equation}
 \bar {\Sigma} (R)  =  \frac{2}{R^2} \int_0^R  \Sigma(R^{\prime}) 
R^{\prime} d R^{\prime}  ~.
\label{sbrav.eq}
 \end{equation}

In the WL regime the  tangential shear $\gamma(R)$  and the excess surface mass 
density  $\Delta \Sigma(R) \equiv \bar {\Sigma}(R) - \Sigma(R)$ 
are  related by  \citep{Um20} 

\begin{equation}
\Delta \Sigma(R) = \gamma(R)  \Sigma_c~,
\label{gamma.eq}
 \end{equation}

where  $\Sigma_c$ is the critical surface mass density:

\begin{equation}
\Sigma_c \equiv \frac{c^2} { 4 \pi G } \frac{D_s} {D_d D_{ds}}~,
\label{sigmac.eq}
 \end{equation}

and $D_s$ , $D_d$  , and $D_{ds}$ are the angular diameter distances between
the observer and the source, from observer to lens, and from the lens to the
 source, respectively.

 The observational quantity of interest is the azimuthally averaged reduced 
 tangential shear profile $g_T(R)$. For comparative purposes, it is  more 
 convenient to express  the radial dependency of  $g_T(R)$ in angular 
coordinates, where for our cosmological model $\sim 10^{''}$  corresponds to
$ \sim 80 \kpc$ at  $z=0.87$:

\begin{equation}
g_T(\theta)=\frac{\gamma(\theta) }{ 1 -\kappa(\theta)}~,
\label{gshear.eq}
 \end{equation}
where $\kappa=\Sigma(\theta)/\Sigma_c$ is the WL convergence and
in the WL approximation $\kappa<<1$.

To compare the measured shear lensing profiles against  those 
 extracted from our SIDM merger  simulations,
 we use  a Burkert profile (\ref{rhogins.eq})
  to model the cored DM density  profiles  of the post-collision DM halos.
The surface mass density (\ref{Sigma.eq}) is accordingly defined as 

\begin{equation}
\Sigma_B(R)= 2 r_c \rho_0 \int_0^{\infty} \frac{ dz^{\prime}} { (1+s)(1+s^2)}
\equiv 2 r_c \rho_0  I(u)~,
 \label{SigmaB.eq}
 \end{equation}
where $s^2=z{^\prime}^2+u^2$, $z{^\prime}=z/r_c$ and $u=R/r_c$. 

 An analytical approximation  to the integral $I(u)$ is given by

\begin{equation}
I(u)\simeq \frac{\pi}{4} \frac{1}{1+u^2}~,
 \label{intpr.eq}
 \end{equation}

with an accuracy to within a few percent for $u\simlt0.8 r_{200}/r_c\sim4$.
Therefore, $\Sigma_B(R)$ reduces to 

\begin{equation}
\Sigma_B(R)\simeq 2 r_c \rho_0  \frac{\pi}{4} \frac{1}{1+u^2}~,
 \label{SigmaBr.eq}
 \end{equation}
and the excess surface density becomes

 \begin{equation}
 \bar {\Sigma}_{B} (R)   \simeq 2 
 r_c \rho_0 \frac{\pi}{4}   \frac{1}{u^2}  \ln (1+u^2) ~.
\label{sbrx2.eq}
 \end{equation}

For each cluster, we can then calculate the lensing profile 
$g^{Burk}_T(\theta)$ at the observer's epoch  using  the best-fit parameters 
of the corresponding Burkert  profile  that was  used to model the  
post-collision DM density profile.
All of the lensing profiles are consistently normalized according 
to $\Sigma_c \simeq 4050  \msun \pc^{-2}$ \citep{Jee14}.

\subsection{Choice of the observational constraints}
\label{subsec:obsvcon}

A successful simulation of the merging cluster El Gordo should be able to 
 reproduce simultaneously the various observational properties that 
characterize the merger.
We list here the most relevant ones that can be used to  constrain 
 the initial condition parameters of the considered SIDM merger simulations.

We construct the list by considering the following available information
on the merging cluster El Gordo: the estimated masses of the primary and 
secondary ($M^{cls}_{200}$), the observed X-ray morphology (\Xmph) and luminosity 
($L_X$), the mean relative radial velocity ($V_r^s$) along the line of sight  
between the SE and NW cluster, 
the measured shear lensing profiles of the two clusters (\glens), the spatial 
offsets between the different peaks ($d_{pk-pk}$),  and finally the 
projected separation ($\dmcl$) at the observer epoch between the mass 
centroids of the two clusters.
We denote the set of all of these constraints as
$\{ M^{cls}_{200},~ X_{mph}^{obs}, ~L_X,~V_r^s,~d_{pk-pk},~g_T(\theta),~\dmcl \}$.

We defer to the following section  a dedicated discussion on the 
most  appropriate, observationally motivated interval of values 
 for the cluster masses $M^{cls}_{200}$.

Regarding the twin-tailed X-ray morphology, the lack of a 
quantitative metric makes it difficult to assess the quality of the agreement 
between the observed morphology and that reproduced by the simulations.
As a solution to this problem, we adopt a  reference model from
the merger simulations of \citetalias{Valda24}, specifically  
the mock X-ray map at the current epoch of  model Bf\_rc20  
(shown  in  top right panel of Figure 2).

According to the findings of \citetalias{Valda24},  this model is one of 
 the fiducial models that best reproduces  the  observed X-ray morphology of 
 the cluster.
 To assess the fidelity of the  X-ray morphology of our 
merging models we will then compare the X-ray maps 
extracted from the simulations against this reference map. 
A detailed description of the  adopted comparison procedures  is 
provided in Section \ref{sec:opt}.

Another noteworthy  feature that can be compared against simulation results
is the relative radial velocity $V_r^s$ between the two clusters.
From   galaxy-based spectroscopic measurements, 
 \citet{Men12}  report an estimate of $V_r^s=598 \pm 96 \kms$,  
 for the  relative radial velocity  of the SE cluster 
 with respect to the NW component.

One of the most significant observational features of the merging cluster 
El Gordo are the spatial offsets between the peak locations of the different 
mass components.
As previously mentioned in the Introduction, the most interesting of these 
offsets is the spatial separation between the X-ray emission peak of the SE 
cluster and the corresponding DM centroid. This is because the offsets is 
positive, that is, the X-ray peak is located farther from the system 
center of mass than the SE DM mass centroid. This is at variance with what is
expected in the commonly accepted outgoing scenario, where the two clusters 
have passed the pericenter and are now observed moving away from each other.

We estimate the uncertainty of the observed offset $d^{SE}_{X-DM}$ as follows.
The size of the positional error for the X-ray emission peak of the SE cluster
is expected to be  relatively small,  set by
the angular resolution ($\sim 0.5^{''}$) of \textit{Chandra}.
For the SE mass peak  the error in the position is given by the WL 
 $\sim 1 \sigma$  error: $\sigma_{DM}\sim 40 \kpc$  \citep{Kim21}.
These estimates allow us to place the offset  in the range 
$d^{SE}_{X-DM} \sim 100 \pm 40 \kpc $.

The positional errors of the other spatial offsets (see  Figure 6 of 
\citet{Kim21}) are insufficient  to discriminate between different merger 
models.  Specifically, the galaxy number density peak of the NW cluster
is  spatially offset by $\sim 120 \kpc$  from its parent mass centroid.
While its positional error is currently unknown, a rough estimate  is  
   $\sim 160 \kpc$ \citepalias{Valda24}.
Moreover, the BCG-DM offset ($\sim 60 \kpc$) of the SE cluster  is 
approximately within the $\sim 1 \sigma$   mass centroid uncertainty 
$\sigma_{DM}\sim 40 \kpc$. 
Finally,  the SZ emission peak is offset from the NW mass centroid by 
$d^{NW}_{SZ-DM} \sim 150 \kpc$. However,  the positional error of the 
SZ centroid is estimated to be relatively large 
($\sigma_{SZ} \sim 70 \kpc \sim 1^{\prime}.4/(S/N)$; \citet{Zh15}).
To summarize, we will consider the offset $d^{SE}_{X-DM} $ as the only useful
 peak-to-peak offset $d_{pk-pk}$ to be included in our list of constraints.

 For the measured shear lensing profiles \glens~of the NW and SE clusters, we 
 adopt the binned data from \citet{Jee14} ( their Figure 9:  \glJ) and \citet{Kim21} 
 (their Figure 6: \glK). To place constraints on our SIDM merger models, these measured
 profiles will be compared against those derived by the DM radial density 
 profiles of the post-collision clusters.

The projected separation $d_{DM}$ between the mass centroids of the two clusters
evolves with time and is  related to the time since pericenter ($t_p$),
which is the elapsed time since the DM mass peaks of the clusters 
reached the pericenter and were at a minimum separation.
We denote the time $t_p$ at the present epoch as $\tpcl$ , and we accordingly 
define $\dmcl$  as the corresponding  value of the projected separation.
The latter   is one of the   most important quantities 
that determine the observational features of the merging cluster El Gordo.
Note  also that, at a given epoch $t_p$,  the value of  $d_{DM}$  
depends on the choice of the projection angle $i$. 

Results of previous SIDM merger simulations \citepalias{Valda24} showed 
that, for a given set of initial conditions,  some observational 
properties   ($X_{mph}^{obs}, ~V_r^s,~d^{SE}_{X-DM}$)
of the merging cluster exhibit a  strong dependency on  $d_{DM}$ and, in turn, on $t_p$.
Observational estimates of the mean projected separation $\dmcl$ range from 
 $\sim 700 \kpc$ \citep{Jee14} up to $\sim 770 \kpc$ \citep{Kim21}, 
 while \citet[][Section 3.1.3]{Ng15}   reported a median value of 
 $\dmcl=740 \pm 7 \kpc$. 
Here we will set $700 \kpc \simlt {\it d^{obs}_{DM}} \simlt 770 \kpc$ 
as the allowed range for $\dmcl$.

\subsection{Preferred range of values for the initial merging dynamical 
parameters and model notation}
\label{subsec:mrgmodels}
This  paper aims to constrain the initial conditions
 parameter space of our SIDM merger models by identifying the subgroup that is able to
 reproduce  the full list of observational constraints
 at the observer epoch: 
$\{ M^{cls}_{200},~ X_{mph}^{obs}, ~L_X,~V_r^s,~d^{SE}_{X-DM},\linebreak 
~g_T(\theta),~d_{DM} \}$.
While a full exploration of the whole parameter space is computationally 
daunting, several simplifying assumptions can significantly constrain the 
allowed range for the  initial merging dynamical parameters 
\{ $\mnw,~\mse,~ V, ~P\}$.

The mass of the El Gordo cluster is a fundamental quantity that determines 
the dynamical evolution of the merging cluster. This has been estimated using 
various methods by a number of authors 
\citep[see, e.g., Table 1 of ][]{Die23}.

These estimates are subject to an uncertainty of a factor of $\sim 2$, 
with  upper limits that  can be as high as $\sim 2.5 \cdot 10^{15} \msun$ 
\citep{Men12,Zi13,Jee14}. At the opposite end, \citet{Die20} infer a total 
cluster mass of  $\sim  1.3 \cdot 10^{15} \msun$. Note that 
all cluster masses here refer  to the definition in Equation (\ref{mcl.eq}).

Given these uncertainties, we will adopt the  lensing-based mass 
measurements of \citet{Jee14} and \citet{Kim21}. According to these authors, 
 the mass of the primary is estimated to be between 
 $\mnw  \sim 1.4 \cdot 10^{15} \msun $ 
\citep[Table 2 of ][]{Jee14} and $\mnw  \sim  10^{15} \msun $ 
\citep[Table 2 of ][]{Kim21}. The mass of the secondary stays close to 
 $\mse \sim 6.5 \cdot 10^{14} \msun $.
Table  \ref{clparam.tab} lists the initial cluster masses 
for the various merger models considered in this work. The first entry 
(model Bf with  $\mnw  = 1.6\cdot 10^{15} \msun $ ) is included for comparison with 
previous runs \citepalias{Valda24}.

It is worth noting that in Table  \ref{clparam.tab}, the masses of the primary 
($\mnw$) of the different models are smaller by about $\sim 30\%$ than the 
chosen limits. This choice is motivated by the accretion effects taking place 
during the  collision between the two clusters, where the primary 
(being the most massive) gains mass from the DM located in the outskirts of the secondary.
According to this effect, the final halo mass  $M^{NW}_{200}$  can be up to  $\sim 30\%$
higher than  its initial value, with  the difference  depending on the collision 
parameters \{ $P, ~V\}$ as well as the observer epoch $t_p$.

 The initial merger configuration is then completed by choosing 
 the appropriate values for the initial relative velocity $V$ 
and  impact parameter $P$. For the chosen range of cluster masses, previous 
findings \citepalias{Valda24} showed that in a standard CDM scenario  
the observed twin-tailed X-ray 
morphology can be reproduced fairly well by merger simulations with  collision
velocities  and impact parameters restricted to the ranges 
 $2,000 \kms  \simlt V \simlt 2,500 \kms$  and 
 $600 \kpc \simlt P \simlt 800 \kpc$, respectively.

\begin{figure*}[!ht]
\centering
\includegraphics[width=0.95\textwidth]{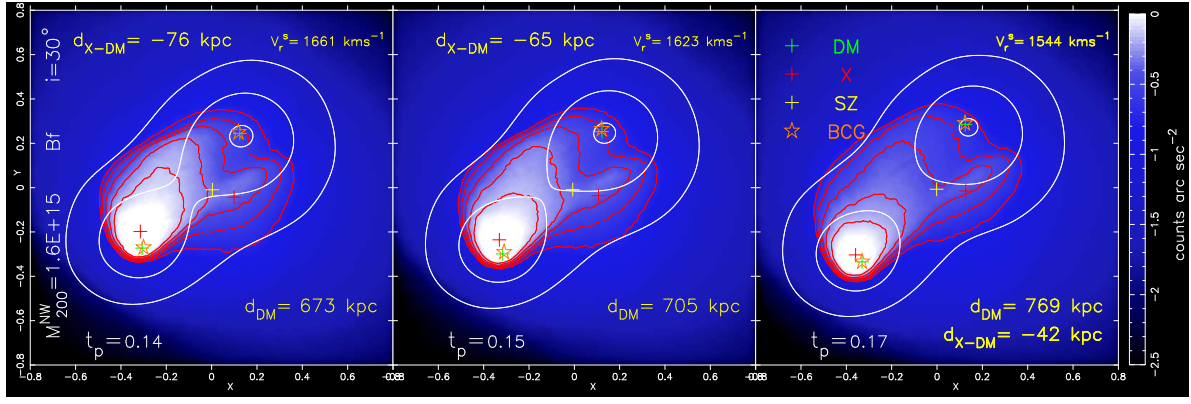}
\caption{ X-ray surface brightness images extracted at various epochs from 
a collisionless  CDM simulation of the merging model 
   ID$_{coll}$-$\{ \sigma_{11},~\sigma_{12},~V_{th} \}$- ID$_{gas}$=
 B\_f-$\{0\}$-Br\_a.
The viewing angle relative  to the merging plane is set to $i= 30 ^{\circ}$, 
 and the  box size is $1.6  \mpc$. Time is in Gyr, with $t_p=0$ corresponding
to the pericenter passage.  
Each panel  reports the projected distance, $d_{DM}$, between the DM 
mass centroids of the two clusters; the relative mean radial velocity 
along the line of sight, $V^s_r$, between the SE and NW  BCG components;
 and the projected offset,  $d_{\XDM}$,  
between the position of the X-ray emission peak and that of the SE DM mass
centroid. A  negative value  for $d_{\XDM}$ indicates  that the  X-ray peak
position is closer to the center of mass of the merging system than  
the mass centroid.
The plus signs  denote the projected spatial locations of 
different centroids : DM mass (green),  X-ray surface brightness (red) and
SZ (yellow).  The projected positions of the BCG mass centroids 
are shown by open orange stars.
Contour levels of the projected X-ray surface brightness (red) and of 
the surface mass density (white) are from the inside to outside at values 
of $(6.6,4.4,2.9,1.9,1.2)\cdot 10^{-1} \ctsn $ and 
$(5.6, 3.1,1.8) \cdot 10^{-1} \dms$, respectively. 
The merger model B\_f-$\{0\}$-Br\_a presented here corresponds to  model 
Bf\_rc20 shown in the top-right map of Figure 2  of 
\citetalias{Valda24}. As outlined in Section \ref{subsec:obsvcon}, 
for consistency purposes, we adopt the  same thresholds and spacing 
for the contour levels of the projected maps	
as those used in previous runs \citepalias{Valda24}. 
The map in the middle panel, at $t_p=0.15$,  can then be directly compared 
with the corresponding top-right map  shown in Figure 2  of 
\citetalias{Valda24}, as the values  of $d_{DM}$ are  approximately the same
and the orientation along the viewing direction  of 
the merging plane is identical.
\label{fig:planeA1}
}
\end{figure*}

 We now make the simplifying assumptions that such constraints will still be 
 valid
  for the SIDM merger models presented here.
  This approximation is based on   the reasonable assumption 
  that  the adopted DM scattering cross-sections  in our  SIDM merger models    
  will not dramatically modify the overall dynamical behavior of the mergers.
  The validity of this approximation will be demonstrated 
  by the simulation results presented in Section \ref{sec:results}.

The  initial collision parameters $\{P, ~V\}$ for the various SIDM merger 
models 
that we investigate are then listed in Table \ref{clparam.tab}.
The table  also reports the halo concentration parameters $\concI$ for the 
primary, as obtained by Equation (\ref{cfit.eq}).
These values are  consistent  at the $1\sigma$ level 
with the  posteriors $\concI(K)=2.54^{+0.9}_{-0.41}$ 
that  \citet{Kim21} report for the NW cluster in their Table 2.
 %For  the halo concentration parameter of the secondary, Equation (\ref{cfit.eq}) would 
% yield  $\concII\sim2.68$ for the chosen  settings of  $\mse $.
 For the chosen  settings of  $\mse $, Equation (\ref{cfit.eq}) yields
 $\concII\sim2.68$ for  the halo concentration parameter of the secondary.
 Such a value is marginally consistent with the posteriors 
 $\concII(K)=3.2^{+1.17}_{-0.74}$ obtained by  \citet{Kim21} 
 for the concentration parameter of the SE cluster.
  Unless otherwise specified,  we will therefore adopt $\concII=3.2$ as the  
  standard value for the concentration parameter of the secondary 
  in our merger simulations.

  Finally, to distinguish the various merger models presented here, 
  we label individual simulations by combining the different labels 
   previously introduced; a generic merging simulation  is thus 
    denoted as ID$_{coll}$-$\{ \sigma_{11},~\sigma_{12},~V_{th} \}$- ID$_{gas}$.
   For example, model 
   Bl-$\{6,4,16\}$-Br\_d  represents a  merger model whose  initial merging 
   dynamical parameters \{ $\mnw,~\mse,~ V, ~P\}$ are given, in the units of
   Table \ref{clparam.tab},
by \{ $10^{15} ,~6.5\cdot10^{14},~ 2000, ~600\}$. According to the notation 
  introduced in Section \ref{subsec:icsidm}, the SIDM parameters  for this
  model are $\sigma_{11}=6\sxu$, $\sigma_{12}=4\sxu $ and $ V_{th}=1600 \kms$.
  Finally, the gas density profiles of the two halos are specified 
  by the entry Br\_d in  Table \ref{subcl.tab}.
  To avoid excessive crowding in the figures, we  identify the merger models 
  in the various panels without including the gas density label ID$_{gas}$;
  however, this is always  made explicit in the corresponding figure captions.

%%%%%%%%%%%%%%%%%%%%%%%%%%%%%%%%%%%%%%%%%%%%%%%%%%%%%%%%%%%%%%%%%%%%

%
\section{Results}
\label{sec:results}

In this  section,  we  present and discuss the main results  
obtained from the various  SIDM cluster merger simulations performed  in this study.
 In Section \ref{sec:opt} we first examine the dependency of several
 observational quantities on the chosen simulation parameters.
From our ensemble  of SIDM simulations, we then present  in 
Section \ref{sec:sidm} the simulation results  of the models that best 
reproduce the main observational properties  of  the cluster.
These  simulations can be  considered our fiducial SIDM merger models, as their
 initial condition parameters   provide the best match  
 to the set of  observational constraints, defined in 
 Section \ref{subsec:obsvcon}, that  characterize  the merging cluster El Gordo.

\subsection{ Analyzing the dependencies of observational properties  on 
simulation parameters}
\label{sec:opt}

In this subsection, we examine how the choice  of various simulation 
parameters  impacts the simulation results used to confront  the 
observational constraints, which are  derived  from the  physical properties
of  El Gordo at the current epoch.
We first  investigate how the mock X-ray maps,  extracted from  the merger
simulations, can vary depending  on  the value of  
 $d_{DM}$   selected  to match  $\dmcl$.

 To this end, we show in Figure \ref{fig:planeA1}  the X-ray images extracted
 at three different epochs $t_p$ from the standard CDM run 
 ID$_{coll}$-$\{ \sigma_{11},~\sigma_{12},~V_{th} \}$- ID$_{gas}$= 
 B\_f-$\{0\}$-Br\_a. Note that, as expected in a standard CDM scenario, the 
 offset $d_{\XDM}$  between the positions of the X-ray  peak and the SE DM mass
centroid  is  negative , that is,  the  X-ray peak position lags behind the 
mass centroid.

 This merger model is identical to  model Bf\_rc20 in \citetalias{Valda24}, 
  whose X-ray map is depicted in the top-right map of Figure 2  of that 
   paper.
 For a consistent comparison, the inclination angle $i$  
along the viewing direction  of the merging plane  has been set 
to the same value ($i= 30 ^{\circ}$). 
As demonstrated by \citetalias{Valda24}, the mock X-ray map extracted from 
this model provides the best match to the
 characteristic twin-tailed X-ray morphology of El Gordo.
  Furthermore,  Figure 6 of \citetalias{Valda24} shows 
that for this merging model the X-ray surface brightness profile, as
measured across the wake, is in accord with {\it Chandra} observations.

 A direct comparison between the X-ray maps in Figure \ref{fig:planeA1}  and 
 the reference map  of  Figure 2  of \citetalias{Valda24}, can be  made 
using the middle panel, at $t_p=0.15$ Gyr. In this snapshot  the value of 
$d_{DM}$ is  approximately the same ($d_{DM}\sim 705 \kpc$) as that reported 
in the reference map.

This map reveals that the twin-tailed 
X-ray morphology is approximately symmetric  around the X-ray emission peak 
of the SE cluster. Moreover, a wake-like  structure is visible behind the peak.
 The  third contour level of the  X-ray emission  (moving outward from the 
 emission peak)  follows this wake and exhibits  a minimum 
located  approximately $\sim 100-200 \kpc$ from the DM centroid of
the NW cluster.
Accordingly, we require the X-ray maps extracted from our SIDM merger 
simulations to reproduce the morphology of our reference map, with a 
specific focus on the behavior of the third contour level.
To ensure a consistent comparison,  we adopt for the contour levels 
 the  same thresholds and spacing  as those previously used  
for the   projected maps  of \citetalias{Valda24}.

 The X-ray maps shown in Figure \ref{fig:planeA1}  
 illustrate how, for a specified set of initial conditions and viewing angle, 
 the projected separation $d_{DM}$, and consequently the X-ray morphology
evolve with the time since pericenter, $t_p$. 

\begin{figure*}[!ht]
\centering
\includegraphics[width=0.95\textwidth]{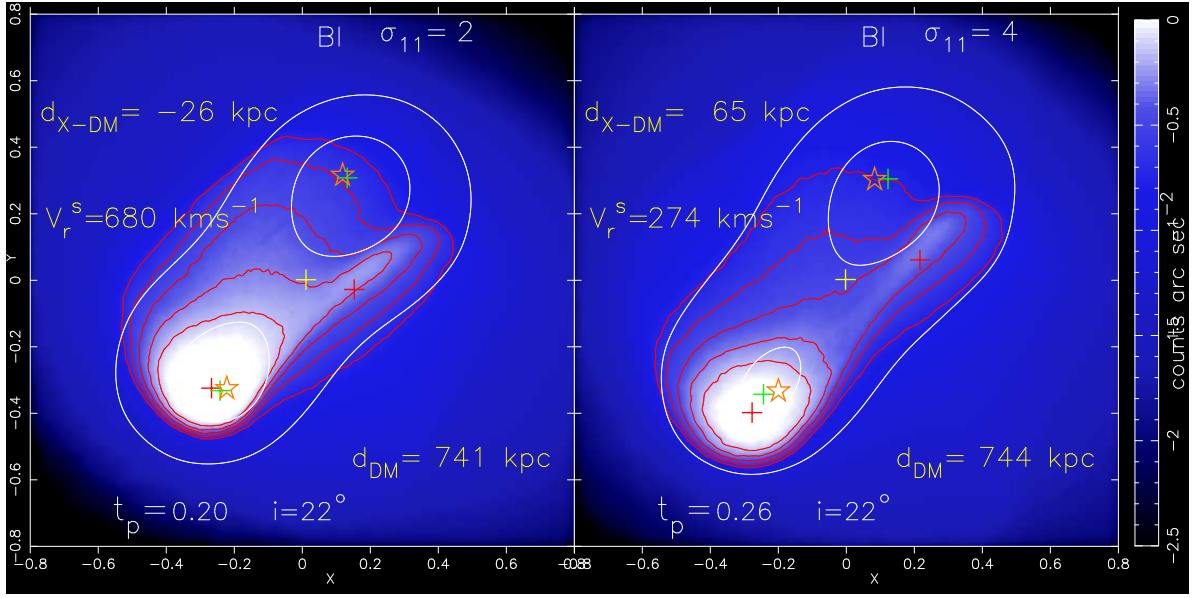}
\caption{ Same as Figure \ref{fig:planeA1}, but for the 
elastic SIDM simulations of merger models 
   ID$_{coll}$-$\{ \sigma_{11},~\sigma_{12},~V_{th} \}$- ID$_{gas}$=
B\_l-$\{2,0,0\}$-Br\_c and B\_l-$\{4,0,0\}$-Br\_c. 
The only difference in the initial settings
of these two simulations is therefore the value of the elastic DM 
cross section: $\sigma_{11}=2 \sxu $ and $\sigma_{11}=4 \sxu $, respectively.
The orientation angle of the merging plane is set to $i= 22 ^{\circ}$.
 In each panel the observer epoch $t_p$ is selected such that 
the projected distance $d_{DM}$  between the DM centroids is approximately 
the same for both models ($d_{DM}\sim 740 \kpc$).
 The meaning of the remaining symbols is the same as in 
Figure \ref{fig:planeA1}.
\label{fig:planeB1}
}
\end{figure*}

As expected, $d_{DM}$ grows with  $t_p$ reaching 
$d_{DM} \sim 770 \kpc$ at $t_p\sim 0.17$ Gyr.  Simultaneously,  the two wings 
of the 
cometary X-ray structure evolve, with the emission from the left 
wing  becoming progressively  weaker. As a result, as  $d_{DM}$ increases, 
the twin-tailed X-ray emission becomes asymmetric around the X-ray peak
and more skewed toward the right.

As previously discussed \citepalias{Valda24}, this behavior is related to the 
different physical processes that occur during the cluster collision, leading 
to the formation of the  observed twin-tailed X-ray morphology.
During the infall phase, the secondary  approaches the pericenter as it 
moves through the  primary's ICM.
 During this phase,  the ICM  is compressed and undergoes a significant 
 increase in temperature and luminosity. 
Subsequently, as the secondary moves toward the apocenter,  
the lower gas temperature and corresponding decrease 
in the ICM sound speed in the primary's outer regions lead 
to the formation of a   bow shock ahead of the secondary 
\citep{Ricker98,Mas08,Mac13,Molnar15,Sh19,Mou21,Cha22}.
This compression of the primary's ICM is responsible for the X-ray emission 
 appearing  as the right wing  around the X-ray peak  seen in the X-ray maps of 
 Figure \ref{fig:planeA1}.

At the same time, as the secondary is falling in the 
potential well of the primary, it will experience ram-pressure effects that 
will strip material from its outer layers, forming  a prominent downstream 
tail. Later, the tail will become misaligned  with the secondary's direction 
of motion, as the cluster  undergoes an off-center collision and  its 
orbit will deviate from the initial trajectory. 
 In the X-ray maps of Figure \ref{fig:planeA1}, this downstream tail 
 corresponds to the left wing of the X-ray emission 
  surrounding  the X-ray peak  (see also Figure 4 of \citet{Zh15}).

%  Scheardown 19 Sect 3. tail  , bow shock pg 10 
%  Chadayammuri 3.1.1 Ricker 98 pg 680 bow shock
The development of an asymmetric X-ray morphology  as 
 the secondary recedes from the primary and $d_{DM}$  increases 
  is a direct consequence  of the distinct physical processes   that 
  generate the observed twin-tailed structure.
 Specifically, while the X-ray emission of the right wing 
  originates by the compression of the ICM  as the 
  secondary moves toward the apocenter,  the left wing is generated 
  by the downstream stripped tail of the secondary.
  In the primary's outskirts,  this material  cools adiabatically 
  and its X-ray emission declines, which in turn produces the observed 
  asymmetry at later epochs.

 %HEREBf21
  These findings are in accord with previous studies
 \citep{Ricker01,Poole06,Sh19};  moreover,  such a behavior is expected to 
 be further exacerbated in merger models featuring  either  a lower-mass 
 primary or self-interacting DM. In both cases, the gravitational potential 
 well of the primary will be less deep or shallower compared to the cluster 
 merger  simulation shown in Figure \ref{fig:planeA1}.
 Consequently,  the post-pericenter gas structures are  expected to be 
 significantly less resilient and more susceptible  to dispersal from 
the  primary's potential well.

 Specifically,  we argue from the current discussion that 
 the symmetric elongation of the two wings observed  around the SE  X-ray peak 
 indicates that  the NW and SE colliding clusters  are currently observed 
shortly after pericenter passage. In particular, the elapsed time since
pericenter cannot exceed $t_p \simlt \tau_X \sim 0.20 \textendash 0.25$ Gyr;
 otherwise, the observed X-ray structure would be significantly more 
 asymmetric (see also Figure 1c in  \citet{Zh15}).

 The temporal evolution of the post-pericenter  X-ray morphology 
 in our merger simulations, 
 when contrasted against the observed twin-tailed X-ray structure,   thus
 provides a powerful diagnostic for deriving significant constraints on the 
 SIDM  parameters required  to reproduce the physical properties of 
 El Gordo.

Finally,  each panel  reports the distance between the SE  DM centroid 
and the X-ray peak ($d_{\XDM}$), together with the relative mean radial 
velocity, $V^s_r$,
along the line of sight between the  two  BCGs. The velocities   are found to 
be of the order of $V^s_r\sim 1,500 \kms$ , in accordance  with the relative
DM bulk velocity 
 between the two clusters at the same epoch.
 This demonstrates  that in a collisionless  CDM merger the velocity distribution 
   of DM particles can be considered a fair proxy of that observed for 
   galaxies \citepalias{Valda24}.
  Moreover,  these collisionless CDM runs consistently exhibit negative offsets 
  $d_{\XDM}$; that is, the post-collision position of the SE DM peak is farther from
  the system center of mass  than the X-ray peak.

\begin{figure*}[!ht]
\centering
\includegraphics[width=0.95\textwidth]{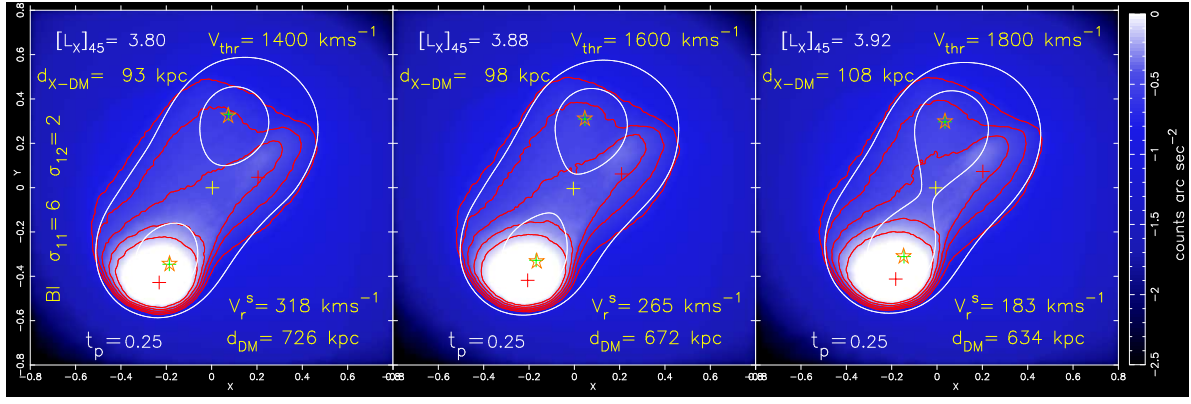}
\caption{  X-ray maps are shown at the same epoch
($t_p=0.25$ Gyr) and  viewing direction ($i= 22 ^{\circ}$)  for 
the inelastic SIDM   merger models B\_l-$\{6,2,14\}$-Br\_d,
 B\_l-$\{6,2,16\}$-Br\_d and B\_l-$\{6,2,18\}$-Br\_d 
(from  left to right).
The cross-sections of the elastic  and  inelastic upscattering 
channels are denoted by $\sigma_{11}$ and $\sigma_{12}$. 
The threshold velocity $V_{th}$ for inelastic scatterings is defined in 
Equation (\ref{thrvb.eq}).  For these models,  the SIDM parameters are 
$\sigma_{11}=6\sxu$ and $\sigma_{12}=2\sxu $,  while 
the threshold velocity varies as 
 $ V_{th}=1,400,~1,600 $ and $ 1,800 \kms$, respectively.
 In each panel, the X-ray luminosity in the $0.5-2$ keV band  is given in 
units of $10^{45} \ergs$.
\label{fig:planeD2}
}
\end{figure*}

The final observational properties of our SIDM merger simulations  clearly depend
 on the value adopted for the elastic DM cross-section $\sigma_{11}$.
To assess how these  properties  vary as a function of $\sigma_{11}$, we  perform two 
distinct elastic SIDM simulations.  To disentangle this dependency from other initial 
conditions, we refrain from  considering inelastic channels in these runs  and keep 
  all  other  initial condition parameters constant.

  We thus considered the two SIDM models, with the parameters set to  
   ID$_{coll}$-$\{ \sigma_{11},~\sigma_{12},~V_{th} \}$- ID$_{gas}$=
B\_l-$\{2,0,0\}$-Br\_c and B\_l-$\{4,0,0\}$-Br\_c, respectively.
Figure \ref{fig:planeB1} shows the mock X-ray maps of these SIDM merger simulations, 
extracted for a viewing  angle of $i= 22 ^{\circ}$ at an observation epoch  
arbitrarily chosen  such that  $\dmcl \sim 740\kpc$. 

From the maps in Figure \ref{fig:planeB1}, it can be seen that both 
  $V^s_r$  and $d_{\XDM}$ clearly exhibit the expected  dependency on 
  the value of the elastic DM cross-section, $\sigma_{11}$.
  This behavior is a specific signature of merging clusters in an SIDM framework.
  During the collision, the DM halos of the two clusters  begin to decelerate 
  as a consequence of the  energy exchange resulting from the scatterings 
  between the DM particles.  After the pericenter passage,  this deceleration 
  manifests as a drag force that acts to  reduce the BCG bulk velocities, 
  thereby decreasing the relative mean radial  velocity $V^s_r$ between the 
  two  BCG components. 
Similarly, depending on the value of $\sigma_{11}$, after the first pericenter
passage positive $d_{\XDM}$  offsets are expected  in SIDM mergers.

These features  are clearly visible 
in the maps of Figure \ref{fig:planeB1}, where the velocities $V^s_r$  are
significantly lower than the values reported in the maps of 
Figure \ref{fig:planeA1}.  Specifically, $V^s_r $ decreases from 
$V^s_r \sim 1,500 \kms $ in the collisionless case 
(see Figure \ref{fig:planeA1}) to  $\sim 680 \kms $ for $\sigma_{11} =2 \sxu$  
 and further down to $\sim 275 \kms $ for $\sigma_{11} =4 \sxu$. 
Simultaneously, the  $d_{\XDM}$  offset between the X-ray  peak and 
the SE  mass centroid is found to increase from 
 $d_{\XDM} \sim -25 \kpc$   to  $d_{\XDM} \sim 65 \kpc$.

These findings illustrate a critical point regarding the limitations
of any SIDM-based approach aimed at reproducing the observational features of 
El Gordo. 
Specifically, if a given SIDM merger model falls short of the observational
range for $V^s_r $ and $d_{\XDM}$,  increasing the elastic
cross section $\sigma_{11}$ helps reduce these discrepancies; 
however,  it simultaneously  decreases the relative bulk motion between
the DM halos and increases the time since pericenter $t_p$ 
(as seen in Figure \ref{fig:planeB1}).
Based on our previous discussion on the X-ray morphology evolution,  
this  implies that at later epochs the system  develops a strongly 
asymmetric X-ray structure around the SE cluster.

To summarize, for any generic SIDM merger model with a given set of
 dynamical parameters  $ \{M_1, ~M_2,~ P, ~V\}$ and halo
gas density profiles,  the range of viable values for the 
elastic cross section $\sigma_{11}$  is expected to be 
constrained  by two competing 
effects.  A small  cross-section ($\sigma_{11} \simlt 2 \sxu$) yields 
a relative radial velocity $V^s_r $ that is too high and an offset 
$d_{\XDM}$ that is too small. Conversely, a cross-section that is too 
high ($\sigma_{11} \simgt 6 \sxu$) leads to an excessively asymmetric 
 X-ray morphology.

 The impact of  an inelastic up-scattering channel, 
 in addition to standard elastic DM self-scattering,  is expected
 to be a reduction in the kinetic energy available after 
 each scattering event. The likelihood of such endothermic reactions 
 exhibits a two-parameter dependency, determined  by both the up-scattering 
 cross section, $\sigma_{12} $, and the velocity threshold $V_{th}$.

 We postulate a range for the inelastic cross section  of the same
 order as  $\sigma_{11} $; otherwise, up-scattering events  
  either would be subdominant ($\sigma_{12} \ll \sigma_{11}$)  or 
 would entirely erase any significant feature associated with elastic 
 scatterings ($\sigma_{12} \gg \sigma_{11}$), such as  the $d_{\XDM}$ offset.

 According to Equation \ref{thrvb.eq}, the velocity threshold $V_{th}$ can be 
 viewed as a high pass filter, where  up-scattering between two DM particles 
 occurs only when their relative velocity $v$ exceeds $ 2 V_{th}$.
 Consequently, in an SIDM merger simulation where all  the initial 
 parameters are held fixed and only $ V_{th}$ is varied, 
 increasing $ V_{th}$ effectively reduces the frequency of up-scattering 
 events. This, in turn, allows the effects of elastic scatterings to
 dominate the post-collision properties of the merger.

 To illustrate this dependency, we show in Figure \ref{fig:planeD2} 
 mock X-ray maps extracted at the same epoch ($t_p=0.25$ Gyr) and 
 viewing direction  ($i= 22 ^{\circ}$), from the SIDM merger models 
   ID$_{coll}$-$\{ \sigma_{11},~\sigma_{12},~V_{th} \}$- ID$_{gas}$=
 B\_l-$\{6,2,14\}$-Br\_d, B\_l-$\{6,2,16\}$-Br\_d 
 and B\_l-$\{6,2,18\}$-Br\_d, respectively. The only difference in the 
 initial conditions of these simulations is the value of the 
 velocity threshold, which for these three models is 
 $V_{th}=1,400,~1,600 \kms$ and $1,800 \kms$, respectively.

 Increasing  $V_{th}$ is  expected to progressively reduce the frequency of 
 up-scattering events;
 conversely, the impact of elastic scattering  on the  post-collision
 properties of the merger becomes more prominent.
 In fact, by extracting the maps at a constant epoch $t_p$, it can 
 be seen that the projected distance between the DM peaks, $d_{DM}$,
decreases from 
 $d_{DM} \sim 726 \kpc $  to $d_{DM} \sim 634 \kpc $ as 
 $V_{th}$  is increased from $V_{th}=1,400,\kms$ to $1,800 \kms$.
 At the same time, the relative radial velocity $V^s_r $ is found to decrease 
 from $V^s_r \sim 320 \kms $  to $V^s_r \sim 180 \kms $.
 The $d_{\XDM}$ offset accordingly increases, albeit with a weaker
 dependence, from $d_{\XDM} \sim 90 \kpc $  for $V_{th}=1,400,\kms$
 to $d_{\XDM} \sim 110 \kpc $ for $V_{th}=1,800,\kms$.

 As outlined in the Introduction, the angle between the axis  connecting 
 the two cluster centers and the plane of the sky cannot be excessively large;
 otherwise, the distinct morphological features exhibited by the merger would 
 be  poorly defined \citep{Men12}. Similarly, \citet{Ng15} employ polarization
 arguments to  set  an upper bound of $i\simlt 21 ^{\circ}$
 for the projection angle. In this work, we construct mock X-ray maps 
 using  three different values for the projection angle $i$:
$i= 15 ^{\circ}$, $i= 22 ^{\circ}$ and $i= 30 ^{\circ}$,  
 with the largest angle  set to $i= 30 ^{\circ}$ to maintain consistency 
 with previous studies \citep{Zh15,Valda24}.

\begin{figure*}[htbp!]
\centering
\includegraphics[width=0.95\textwidth]{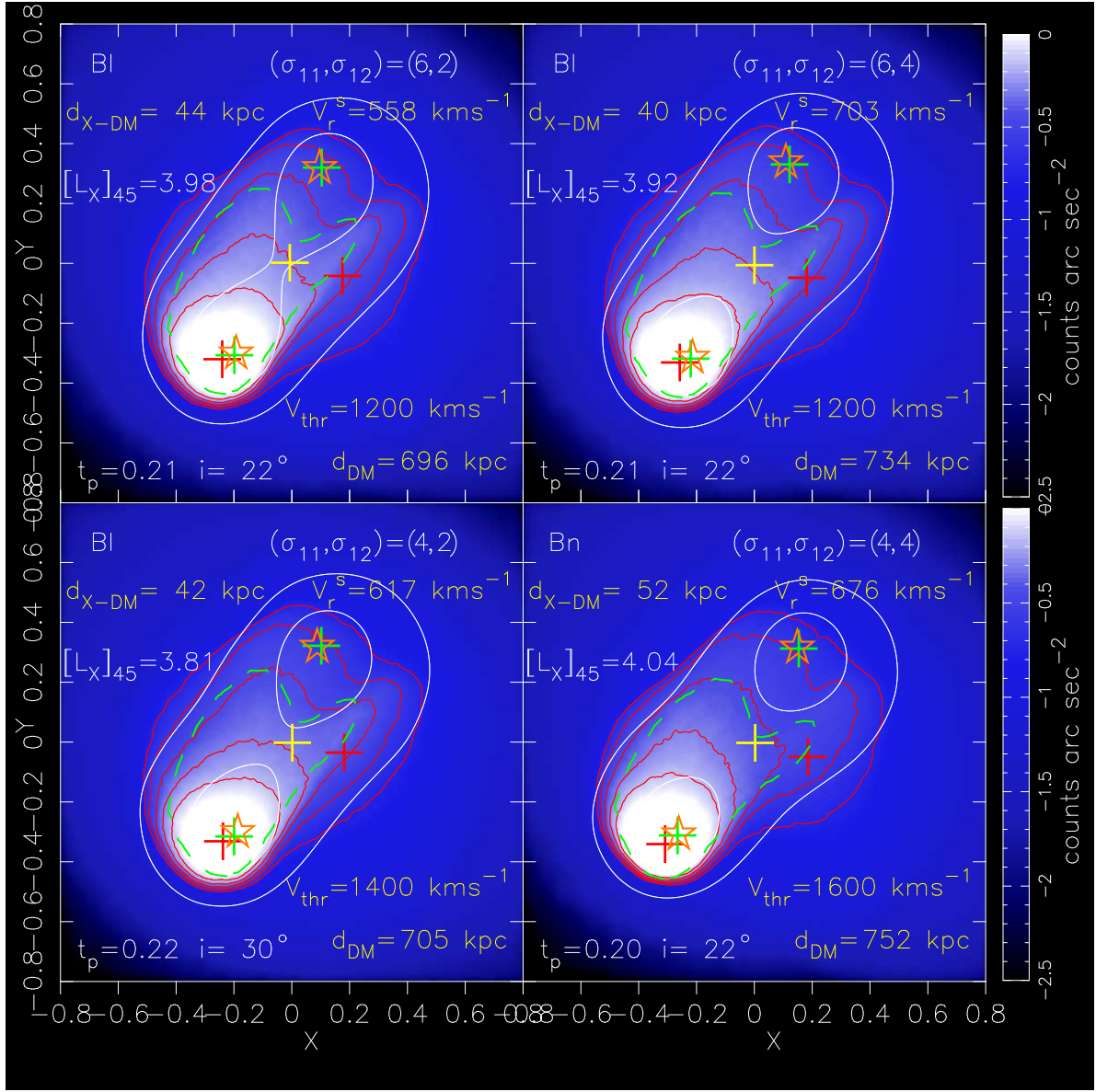}
\caption{ 
Mock X-ray maps extracted at various epochs and 
viewing angles from four fiducial SIDM merger models. These models
were selected because  merging simulations with these    initial conditions 
best reproduce the set of  observed properties   
$\{ X_{mph}^{obs},~L_X,~V_r^s,~d^{SE}_{X-DM},~\dmcl \}$ at the observation 
epoch.
Moving clockwise from the top left panel, these models are 
Bl-$\{6,2,12\}$-Bm\_a, Bl-$\{6,4,12\}$-Bm\_a, Bn-$\{4,4,16\}$-Br\_e and 
Bl-$\{4,2,14\}$-Bm\_a, respectively.
 For all of these merger models, the mass of the primary is set to
$M^{(NW)}_{200}= 10^{15}\msun$ (see Table \ref{clparam.tab}).
Each panel indicates  the  observer epoch $\tpcl$, the viewing angle $i$ and 
the corresponding mean projected separation $d_{DM}$, which ranges  from 
 $\sim 700 $  up to $\sim 750 \kpc$.
 In each panel the dashed green line represents the third contour level,  
 extracted  at the present epoch from the reference map in the middle panel of 
Figure \ref{fig:planeA1}, of the fiducial merger model B\_f-$\{0\}$-Br\_a; 
the positions of the points defining the curve have been normalized to the 
position of the SE X-ray peak (see text).
As in the previous figures the plus signs denote the projected
 spatial positions of the different centroids : DM mass (green),  X-ray 
surface brightness (red) and SZ (yellow). The open orange stars mark
the projected positions of the BCG mass centroids.
\label{fig:planeJ2}
}
\end{figure*}

 For the observables $d_{DM} $  and $V^s_r $,  the expected dependency on the 
 projection angle $i$  follows $d_{DM} \propto \cos(i) $, with  the offset
 $d_{\XDM}$  exhibiting a similar geometrical scaling, while the radial 
 velocity scales as $V^s_r\propto \sin(i)  $  \citep{Greg84}.
 For example, model  B\_l-$\{6,2,14\}$-Br\_d, in the left panel of 
 Figure \ref{fig:planeD2},   has  a separation and a 
 radial velocity that  vary from  $d_{DM} = 754 \kpc $  ($V^s_r =225 \kms $) at  
$i= 15 ^{\circ}$ to $d_{DM} = 701 \kpc $  ($V^s_r = 415 \kms $) at $i= 30 ^{\circ}$. 
From this example, based on the  considered  uncertainties in the projection angle $i$, 
it is possible to deduce  the expected range of variations for these observables.

In summary,  the examples presented in this section  
 aim to quantify the uncertainties inherent in
 drawing conclusions \textendash~ based on current observational constraints
 \textendash~
 about the validity of a specific SIDM merger model.

\subsection{El Gordo fiducial merger models from two-state endothermic SIDM 
simulations  }
\label{sec:sidm}

In this subsection, we present the fiducial SIDM merger models that 
 best match the set of observational constraints for the El Gordo cluster,
$\{ M^{cls}_{200},~ X_{mph}^{obs}, ~L_X,~V_r^s,~d_{pk-pk},~g_T(\theta),~\dmcl \}$, 
as discussed in Section \ref{subsec:obsvcon}.

The initial parameters for these models were identified by first exploring a
broad suite of SIDM merger simulations and subsequently narrowing the parameter
space until the simulations results reached  a satisfactory consistency  
with the specified set of constraints.

Following the discussion of   Section \ref{subsec:mrgmodels}, 
our ensemble of SIDM merger simulations is constructed  by considering 
 two distinct subsets based on the following  mass of the primary: 
 $\mnw  \sim 1.4 \cdot 10^{15} \msun $ \citep{Jee14}, and 
 $\mnw  \sim  10^{15} \msun $ \citep{Kim21}. 
 The first subset comprises   the merger models in Table \ref{clparam.tab}
with ID$_{coll}$ labels Bl and Bn, while  models Bw and By constitute  the 
second subset.

 Note again that in Table \ref{clparam.tab}, the initial  masses of 
 the primary clusters are lower than the targeted final  values.
This  difference accounts  for accretion effects during the merger, 
which \textendash~ at the observer epoch \textendash~
increase the  primary's mass by $\sim 20-30\%$.
Furthermore, it can be seen from Table \ref{clparam.tab} that,
 within the considered range for the primary mass, the initial  impact parameter is set 
 to $P=600 \kpc$ for all  the merger models investigated. 

This assumption is based on previous findings \citepalias{Valda24}, 
which demonstrated  that  merger simulations  can  reproduce the observed 
twin-tailed X-ray morphology only within a  narrow range of initial  impact 
parameters centered around $P\sim600 \kpc$ (see, in particular, Figure 5 of 
\citetalias{Valda24}).
This follows because  as $P$ approaches  zero  the merger becomes a head-on 
collision. Conversely,  for  very large values of $P$ ($\simgt 800 \kpc$),
the ram pressure experienced
by the secondary during the collision is significantly weaker;  this, in turn, 
results in  an asymmetric, one-tailed X-ray structure.  
Similar arguments can be used to demonstrate \citepalias{Valda24} that 
 the appropriate range  for the infall velocity $V$   
is  $ 2,000 \kms \simlt V  \simlt 2,500 \kms$.

\begin{figure*}[htbp!]
\centering
\includegraphics[width=0.95\textwidth]{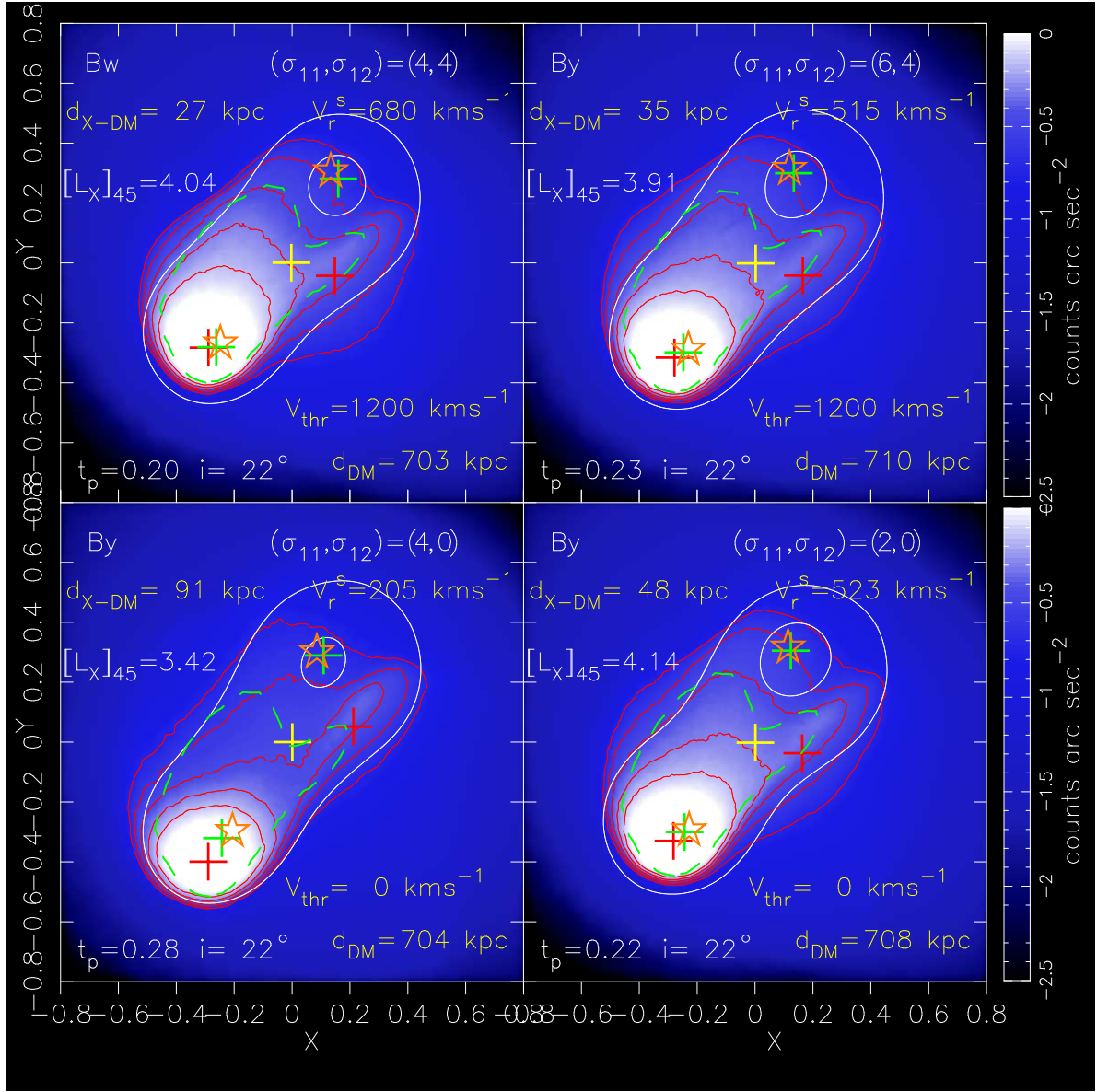}
\caption{Same as in Figure \ref{fig:planeJ2}, but for the merger models 
Bw-$\{4,4,12\}$-Bm\_b (top left panel) and By-$\{6,4,12\}$-Bm\_b 
(top right panel).
 Here the primary's masses  are set to
$M^{(NW)}_{200}= 8.5 \cdot 10^{15}\msun$.
The X-ray maps shown in the top panels correspond  to the
observational epoch $t_p$ at which   the  set of observed properties   
$\{ X_{mph}^{obs},~L_X,~V_r^s,~d^{SE}_{X-DM},~\dmcl \}$ is best reproduced. 
The bottom left and bottom right  panels show maps  of  two elastic SIDM
simulations: models By-$\{4,0,0\}$-Bm\_b and By-$\{2,0,0\}$-Bm\_b, respectively.
The projected separation $d_{DM}$ in these bottom panels is approximately equal
to that of the two merger models displayed in the top panels.
\label{fig:planeJ3}
}
\end{figure*}

 Finally,  Table \ref{subcl.tab} lists the initial parameters for the cluster
gas density profiles employed in our simulations. It is  worth emphasizing that
these are the  parameters used in  the simulations presented here;
the full region of parameter space explored during the   initial testing
of the gas profiles  was approximately  five times larger.

 We show in Figure \ref{fig:planeJ2} the  X-ray surface brightness maps
 for the first subset of four fiducial SIDM merger models, which use the 
 initial 
 collision parameters of models Bl and Bn from  Table \ref{clparam.tab}.
 For these  models, the final mass of the primary is expected to be 
  $M_{NW}\sim 1.3 \cdot 10^{15}$. For each SIDM model, the 
 observer epoch 
 and viewing angle are chosen to best reproduce  the set of  observational 
 properties   
$\{ X_{mph}^{obs},~L_X,~V_r^s,~d^{SE}_{X-DM},~\dmcl \}$.

Moving clockwise from the top left panel, these models are: 
   ID$_{coll}$-$\{ \sigma_{11},~\sigma_{12},~V_{th} \}$- ID$_{gas}$=
Bl-$\{6,2,12\}$-Bm\_a, Bl-$\{6,4,12\}$-Bm\_a, Bn-$\{4,4,16\}$-Br\_e and 
Bl-$\{4,2,14\}$-Bm\_a.
The contour level indicated in each panel by the dashed green line corresponds
to the third contour level shown for the fiducial merger model 
B\_f-$\{0\}$-Br\_a in the mock X-ray map of Figure \ref{fig:planeA1} 
(middle panel).  As explained in  the previous section, we use this map as a 
benchmark  against which the results of our SIDM merger simulations are 
compared.

Specifically, for any SIDM model under consideration, we assess  the 
morphological consistency of its X-ray maps by using the behavior of the 
third contour level to discriminate between different models.
We construct the reference level by first extracting, from the original map 
of Figure \ref{fig:planeA1}, a set of points that defines  the contour level 
of interest; the origin of their $\{\hat x,\hat y\}$ coordinates is placed  
at the position of the SE X-ray peak. These points are then used to generate
the contour levels (dashed green lines) shown in the maps of  
Figure \ref{fig:planeJ2}. In each map, the origin of the $\{\hat x,\hat y\}$ 
coordinate system is likewise centered on the SE X-ray peak; the curve defined
by 
these points is then rigidly rotated until it visually achieves the best 
match with the third contour level of the current X-ray map.

A critical investigation of the maps of Figure \ref{fig:planeJ2} enables us to 
draw several conclusions about the validity of these SIDM merger models.
In all four models the X-ray morphology is   well reproduced, as are 
 the mean radial velocities $V^s_r$, which fall
 within the measured range of $\sim 600 \pm 100 \kms$ \citep{Men12}.

The offsets $d_{\XDM}$   are relatively small ($\sim 40 \kpc$) with respect to
the estimated interval of $d^{SE}_{X-DM} \sim 100 \pm 40 \kpc $. The only 
exception is provided  by model Bn-$\{4,4,16\}$-Br\_e  (bottom right panel), for which 
  $d_{\XDM} \sim 50 \kpc$. This is also the model that exhibits the largest 
   projected separation $d_{DM}$ ($\sim 750 \kpc$). A simple solution to 
  increase $d_{\XDM}$  would be to increase  either the value of 
 $\sigma_{11} $, as shown in the models of   Figure \ref{fig:planeB1}, or 
 the value of $d_{DM}$. However, both  approaches 
would imply a  later  observer epoch $\tpcl$, which, according to the 
discussion of Figure \ref{fig:planeA1}, would in turn produce an 
asymmetric X-ray morphology.

 Another approach would to be to increase the initial collision velocity $V$ of
 the merger model, which, according to  Table \ref{clparam.tab} is 
 $V=2,500 \kms$ for model Bn. However, when the collision velocity $V$ is too 
 high relative to the infall velocity set by the cluster masses (see Section 
 3.1.1 of \citetalias{Valda24}), the duration of the interaction between the 
 two clusters is relatively short and the resulting X-ray morphology is 
 strongly asymmetric (see also Figures 5a and 5b of 
 \citet{Zh15}).

 An interesting feature  shown in the X-ray maps of  Figure \ref{fig:planeJ2} 
 is the absence of significant BCG-DM offsets for the SE cluster. 
 To avoid excessive crowding , these  values are not reported directly on the
 maps in Figure \ref{fig:planeJ2}; however,  for all the merger models, they 
 are found to lie in the range $d_{\BDM} \simlt 10 \kpc $. Such values are 
 significantly smaller than the corresponding offsets ($d_{\BDM} \sim 70 \kpc$) 
 found for  previously proposed SIDM merger models (cf. model XDBf\_sb in 
 Figure 12 of \citetalias{Valda24}),
 and are well within the $1 \sigma$ upper limits set for the SE cluster by the 
 WL mass centroid uncertainty \citep{Kim21}.

 A plausible explanation for the smallness of these values is provided by the 
 presence of an up-scattering channel, which is expected to reduce the impact
 of elastic scatterings. Nonetheless, as shown in Figure \ref{fig:planeB1}, 
  small values of $d_{\BDM}$ are maintained even for the purely elastic SIDM 
  simulation B\_l-$\{4,0,0\}$-Br\_c. 

\begin{figure*}[htbp!]
\centering
\includegraphics[width=0.95\textwidth]{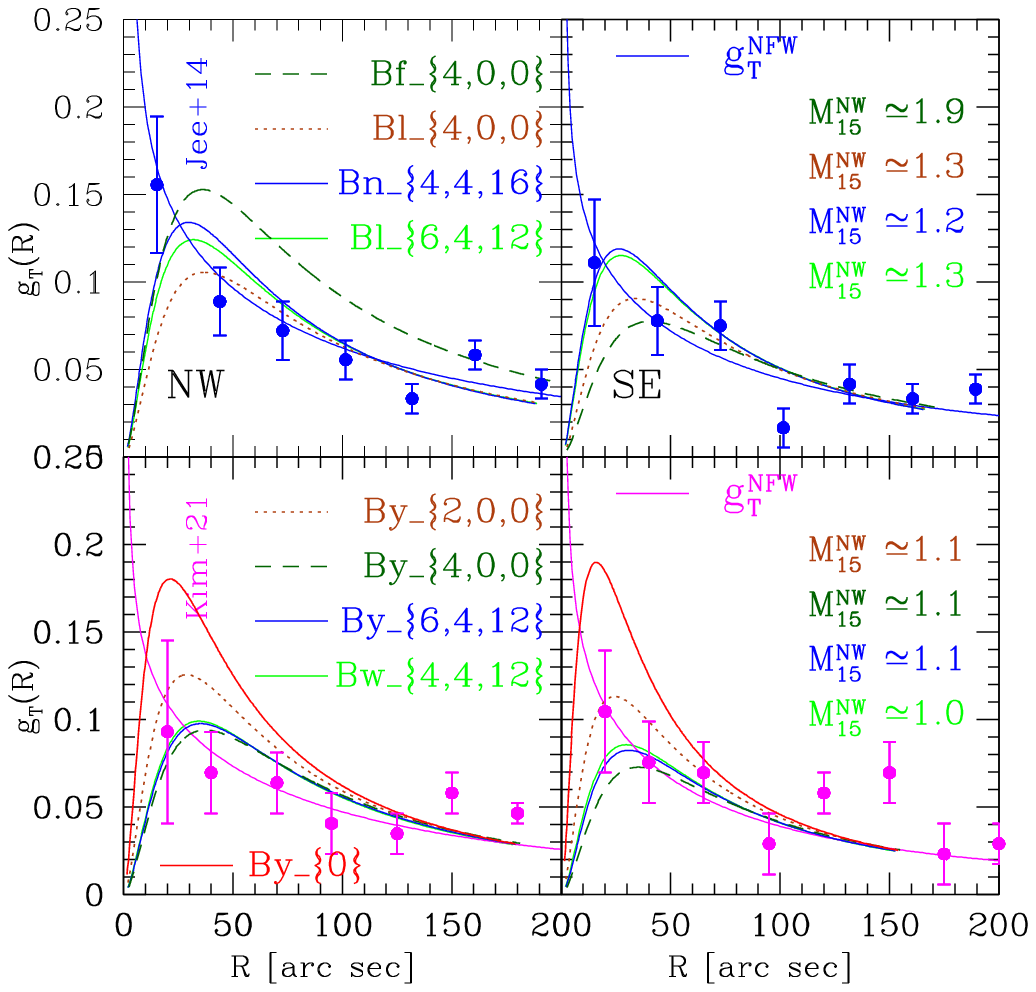}
\caption{
Reduced tangential  shear profiles, $g^{Burk}_T(\theta)$, calculated 
from Burkert density profiles according to Equation (\ref{gshear.eq}).
The left and right panels show the profiles 
 for the  NW and SE clusters, respectively. The top panels display the 
profiles derived from  the cored DM distribution extracted from two of the
fiducial SIDM merger simulations shown in Figure \ref{fig:planeJ2}: models
 Bn-$\{4,4,16\}$-Br\_e and Bl-$\{6,4,12\}$-Bm\_a.
For comparison, we also include the profiles (dark green dashed and 
brown dotted lines) corresponding  to elastic SIDM simulations with initial 
merger parameters  Bl-$\{4,0,0\}$-Bm\_a  and Bf-$\{4,0,0\}$-Br\_b.
 For these  SIDM merger models, 
$M_{15}^{NW}$ in  the top right panel denotes 
the mass $M_{200}$ of the NW cluster in units of $10^{15}\msun$.
The bottom panels show the profiles extracted from the merger models  
Bw-$\{4,4,12\}$-Bm\_b and By-$\{6,4,12\}$-Bm\_b, as presented in the top 
panels of Figure \ref{fig:planeJ3}.
For comparison, the green dashed and brown  dotted  lines correspond to the 
lensing profiles derived from  the two elastic SIDM simulations, 
 models By-$\{4,0,0\}$-Bm\_b and By-$\{2,0,0\}$-Bm\_b, 
which are shown in the  bottom panels of Figure \ref{fig:planeJ3}.
 Additionally,  we also display  the WL profiles 
 derived from a collisionless CDM merger model (solid red lines).  
This reference case is labeled By-$\{0\}$-Bm\_b and represents  the parent simulation
of model By-$\{2,0,0\}$-Bm\_b, but with $\sigma_{11}=0$.
 Observational data points,  extracted from Figure 9 of \citet{Jee14} and 
Figure 17 of \citet{Kim21}, are displayed  in the top and bottom panels, 
respectively.
In each panel, the data points are  contrasted against a
lensing  profile  $g^{NFW}_T(\theta)$, calculated from  a best-fit  NFW 
mass model. 
The $g^{NFW}_T(\theta)$ profiles in the top panels are constructed 
according  to the best-fit NFW parameters taken from Table 2 of \citet{Jee14}: 
$ \{ r_{200}^{NW}, r_{200}^{SE} \}= \{1.65,1.38\} \mpc$.
and $\{ \concI, \concII \} = \{ 2.57, 2.65 \}$.
In the bottom panels  the $g^{NFW}_T(\theta)$  profiles are derived
from  the NFW parameters reported in  Table 2 of \citep{Kim21}:
$ \{ r_{200}^{NW}, r_{200}^{SE} \}= \{1.5,1.3\} \mpc$ 
and $\{ \concI, \concII \} = \{ 2.54, 3.20 \}$.
\label{fig:plensJ2}
}
\end{figure*}

 We then interpret the very small sizes of the offsets $d_{\BDM}$ as a 
 manifestation
 of a reduced deceleration of the DM components during the cluster collision.
 Specifically, the effective drag  acceleration acting on the SE halo will be 
 proportional to  $ \propto \sigma_{11} M_{SE} V_{SE}^2$ 
 \citep{Harvey14,Kah14,Secco18}, where $V_{SE}$  is the halo velocity.  
 For models Bl and Bn the initial mass of the 
NW cluster ($M_{NW}=10^{15}$) is  $\sim 40 \%$ lower than that of 
model XDBf\_sb (model Bf: $M_{NW}\sim 1.6 \cdot 10^{15}$;  see  Figure 12 
of \citetalias{Valda24}). For these models, the infall velocity $V_{SE}$  
at the pericenter is then  $\sim 20 \%$ lower than in model Bf, 
resulting in a drag term  that is  $\sim 40 \%$ smaller.

 Finally, an important   results is that in all  the merger models the 
 final X-ray luminosity $L_X$ is approximately a factor of $\sim 2$   higher than 
 the estimated value of $L_X \sim 2\cdot 10^{45} \ergs$ \citep{Men12}. We 
 attribute  this discrepancy to a direct consequence of the initial cluster 
 gas fractions; as shown in  Table \ref{subcl.tab}, these have been raised to 
 cosmological levels ($ f_g \sim 0.16$) for all the merger models. We will 
 return to this topic in the Conclusion.

We now  discuss the second subset of fiducial SIDM merger models, 
which consists of  models  Bw-$\{4,4,12\}$-Bm\_b and 
By-$\{6,4,12\}$-Bm\_b; for these models the final masses of the primary  
are $\mnw  \sim 10^{15} \msun $. The mock X-ray maps for these 
fiducial models   are shown in the top panels of Figure \ref{fig:planeJ3}.

From these maps, it can be seen that while the X-ray morphology is  
reproduced fairly well for both of the models, the projected separation 
between the DM mass centroids is limited to $d_{DM}\simlt 710 \kpc$.
Consequently, this also implies that the offsets $d_{\XDM}$  are constrained 
to values below $d_{\XDM}\simlt 40 \kpc$. 
Increasing $d_{DM}$, and by consequence $d_{\XDM}$,  
by considering later epochs is not a viable solution.
As  demonstrated in Section \ref{sec:opt}, a significant side effect of this 
approach would be  the introduction of highly  asymmetric X-ray morphologies at the observed epoch.

%For the four SIDM merger models of Figure \ref{fig:planeJ2} the reduced 
%tangential shear profiles $g^{Burk}_T(\theta)$ are shown in Figure 
%\ref{fig:plensJ2}.
In the top panels of Figure  \ref{fig:plensJ2} we show the reduced 
tangential shear profiles, $g^{Burk}_T(\theta)$,  for two of the four 
SIDM merger models of Figure \ref{fig:planeJ2}:
 Bn-$\{4,4,16\}$-Br\_e and Bl-$\{6,4,12\}$-Bm\_a. The lensing profiles 
 of the remaining two Bl models are omitted for brevity, as they are nearly 
 identical to the profiles of the selected  Bl model.

For each  model, the profiles were calculated at the time $t_p$ 
indicated in  the corresponding maps of Figure \ref{fig:planeJ2}. The   
profiles are shown separately for the NW and SE cluster in the left and 
right panels, respectively.
These profiles were obtained according to the procedures described in 
Section \ref{subsec:lensprof},  utilizing a  Burkert  profile  to model the 
post-collision DM density distribution of the halos.

For comparative purposes, we additionally present lensing profiles 
extracted from two elastic SIDM simulations with the initial 
merger parameters  of models Bf-$\{4,0,0\}$-Br\_b and 
Bl-$\{4,0,0\}$-Bm\_a. 
For both of these models, the present epoch is identified when 
 $d_{DM} \sim 700 \kpc$. We note that model Bf-$\{4,0,0\}$-Br\_b is 
 identical to  model XDBf\_sb of \citetalias{Valda24},  and  its corresponding
 lensing profiles were previously presented in Figure 3 of \citet{V25}.

We also include the measured profiles  extracted from Figure 9 of \citet{Jee14} 
(top panels) and Figure 17 of \citet{Kim21} (bottom panels).
Finally, we contrast these profiles against an analytical lensing profile 
$g^{NFW}_T(\theta)$,  derived from a NFW  model \citep{Um20},  used  by those 
authors to describe the cluster mass distribution (see figure caption for 
details). 

For completeness, we indicate the final primary mass, $M_{NW}$, 
of each merger model  in the right panels of 
Figure \ref{fig:plensJ2}; 
 these values are  deduced at the observer epoch from the corresponding 
 post-collision mass distribution.
As expected, for the  two  selected SIDM merger models of Figure 
\ref{fig:planeJ2}, as well as for the 
other two, the final masses of the primary remain close to 
$\mnw  \sim 1.3 \cdot 10^{15} \msun $. Consequently,  a meaningful 
comparison for these models should be made with the binned profiles of 
\citet{Jee14},  shown in the top panels,  as the
estimated  primary mass  was 
$\mnw  \sim 1.4 \cdot 10^{15} \msun $.

We first comment on the profiles $g^{Burk}_T(\theta)$  extracted from the 
two elastic SIDM simulations.  
The lensing  profiles of model Bf-$\{4,0,0\}$-Br\_b  are clearly inconsistent 
with the measured profiles. This is not surprising given that, for this model,
 the final mass of the primary is  $\mnw  \sim 1.9 \cdot 10^{15} \msun $, 
 and large core radii ($ r_c \sim 300 \kpc$)   develop
 in the DM density profiles of the two clusters during the collision  (see 
 also Figure 2 of \cite{V25}).

 For merger model Bl-$\{4,0,0\}$-Bm\_a, there is now a significant 
 improvement with the binned profiles of \citet{Jee14}, as the final mass of 
 the primary is now $\mnw  \sim 1.3 \cdot 10^{15} \msun$.
 Nonetheless, for the innermost bin at $R\sim 20 ^{''}$, the 
 $g^{Burk}_T(\theta)$  profiles of both the NW cluster and the SE cluster still fall
 short of the measured values.

 The situation changes significantly for the selected two fiducial SIDM merger
 models.
 All of the $g^{Burk}_T(\theta)$  profiles of these models show 
 much better agreement with the binned data in the innermost regions, with 
 very little variation between the profiles of the different models
 across the entire angular range.

This represents one of the most significant findings of our study, 
as it clearly demonstrates that within an SIDM framework the presence of 
an up-scattering channel is pivotal; the channel allows for the consistency 
of the merger models with 
lensing data while simultaneously satisfying other observational
constraints.

 The bottom panels of Figure \ref{fig:plensJ2} shows the lensing profiles
 $g^{Burk}_T(\theta)$,  
 as derived  from the two SIDM merger models presented in the top panels of 
 Figure \ref{fig:planeJ3}. For these models  $\mnw  \sim 10^{15} \msun $; 
  a meaningful comparison can thus be made  with the measured profiles 
  depicted in the bottom panels,  which were extracted from 
  Figure 17 of \citet{Kim21}.
As shown, within the observational uncertainties, both  
models exhibit lensing  profiles in good agreement with the binned data.
This demonstrates that for these models as well the presence of an 
up-scattering channel is 
fundamental to achieving consistency with the measured lensing profiles.

In order to assess the significance of the endothermic channel, for comparative 
purposes we additionally show lensing profiles extracted from two 
 elastic SIDM simulations, with  initial merger parameters  given by 
 By-$\{4,0,0\}$-Bm\_b  and By-$\{2,0,0\}$-Bm\_b, respectively. 
 These merger simulations were specifically chosen with the same initial 
 conditions as model By-$\{6,4,12\}$-Bm\_b, with the only difference 
 being the choice of the SIDM parameters.
 As in the top right panel of Figure \ref{fig:planeJ3}, the observer epoch  
 for these merger models is consistently identified as the time $t_p$ when
 $d_{DM}\sim 710 \kpc$.

 As seen in the bottom panels of Figure \ref{fig:plensJ2}, at small angles 
 ($R\simlt 50 ^{''}$)
 the lensing profiles derived from model By-$\{2,0,0\}$-Bm\_b  exhibit 
 a significant excess compared to the binned data. This is not 
 the case for model By-$\{4,0,0\}$-Bm\_b, whose profiles follow an angular 
  dependency very similar to that of the inelastic models. 

 It would be tempting to interpret this result as an indication 
 that the observational properties of El Gordo can be explained by 
 a simpler, elastic SIDM scenario. However, the X-ray surface 
 brightness maps extracted from the two elastic models demonstrate that 
 this is not the case.

 In the bottom panels of Figure \ref{fig:planeJ3}, we show  the mock X-ray maps 
 derived from the two models By-$\{4,0,0\}$-Bm\_b  and By-$\{2,0,0\}$-Bm\_b,
 at the same epochs for which the corresponding lensing profiles in 
 Figure \ref{fig:plensJ2} 
 were calculated. For both  models, the X-ray morphologies are strongly 
 asymmetric and clearly inconsistent with the reference contour level
  (green dashed line).  Moreover, for model By-$\{4,0,0\}$-Bm\_b, the relative
  radial velocity $V^s_r $ is found  to be significantly lower than 
  observational estimates ($V^s_r \sim 200 \kms$).

  Furthermore, for comparative purposes, we also show the WL profiles 
  derived from a collisionless CDM merger model  in the bottom panels (solid red lines).
  This merger model is labeled By-$\{0\}$-Bm\_b and represents  the parent 
  simulation of model By-$\{2,0,0\}$-Bm\_b, but with $\sigma_{11}$ set to zero.

  For this model the lensing profiles $g_T(\theta)$  of the NW and SE clusters are 
  consistently extracted from the simulation at the same time $t_p$ when 
  $d_{DM}\sim 710 \kpc$, as in the case of models By-$\{4,0,0\}$-Bm\_b  and 
  By-$\{2,0,0\}$-Bm\_b.
  It can be seen that for this model, at small angles ($\theta \simlt 50^{\prime}$),
  the lensing profiles $g_T(\theta)$  tend to diverge significantly from the binned
  data, thereby highlighting the difficulty of the standard CDM scenario in explaining
  the observational features of El Gordo.

 Finally, it is worth noting in Figure \ref{fig:plensJ2} that at small 
angles
($\theta \simlt 50^{''}$) the lensing profiles $g^{Burk}_T(\theta)$
extracted from the SIDM simulation differ significantly  from the 
NFW  profiles $g^{NFW}_T(\theta)$, which diverge as $\theta \rightarrow 0$.
 This inconsistency is highly significant for all  considered SIDM models 
 and  demonstrates that future WL measurements will serve as a decisive 
 test bench for these scenarios 
(see point ii) of the Conclusions regarding the impact of future WL analysis 
  of El Gordo cluster on SIDM merger models).

  We conclude this section by noting that the two fiducial SIDM merger models
  of Figure \ref{fig:planeJ3}, would clearly encounter difficulties if future 
  WL mass measurements were to establish an observational framework 
  for El Gordo characterized by a low mass of the primary
  (say,  $\mnw  \sim 10^{15} \msun $) and,  a the same time,  
   a large value of the projected  separation
 ($d_{DM} \simgt 750 \kpc$). We will return to this issue in the Conclusions.

\section{Summary and Conclusions}
\label{sec:discuss}
In this work, we present a simulation study of the high-redshift, massive 
 merging cluster El Gordo.
Our ensemble of N-body/hydrodynamical binary merger simulations 
was constructed within a two-state SIDM scenario, exploring a wide range 
of initial conditions.  
These simulations were aimed at identifying  the subset 
of the  initial condition parameter space that  best reproduces 
the set of observational constraints previously defined  in Section 
\ref{subsec:obsvcon}: 
$\{ M^{cls}_{200},~ X_{mph}^{obs}, ~L_X,~V_r^s,~d_{pk-pk},~g_T(\theta),
~\dmcl \}$.

In accordance with WL mass measurements \citep{Jee14,Kim21}, the final mass of
the primary cluster in the simulations was constrained within the range 
$ 10^{15} \msun \lesssim \mnw \lesssim  1.4 \cdot 10^{15} \msun $, 
ensuring  that the first  observational constraint $M^{cls}_{200}$ was
satisfied by definition. Furthermore, we specifically investigate a two-state
inelastic, endothermic SIDM  scenario to explicitly  determine whether such 
framework can produce  merger simulations in agreement  with measured lensing 
profiles.

From the simulation ensemble, we identified six  fiducial SIDM merger models
that provide the best match to the specified  set of observational 
properties  for the El Gordo cluster. The initial  collision velocity $V$  and 
impact parameter $P$ for all of these models lie in the range 
$1,800 \kms \lesssim V  \lesssim 2,500 \kms$ and $P \sim 600 \kpc$, 
respectively. 

For  a mass of the primary around $\mnw  \sim 1.3 \cdot 10^{15} \msun $,  
we identified a first subset of four fiducial SIDM merger models: 
Bl-$\{6,2,12\}$-Bm\_a, Bl-$\{6,4,12\}$-Bm\_a, Bn-$\{4,4,16\}$-Br\_e and 
Bl-$\{4,2,14\}$-Bm\_a. The corresponding mock X-ray  maps 
 for these models are presented in Figure \ref{fig:planeJ2},
 while the corresponding lensing profiles for two of 
  these models  are shown  in Figure \ref{fig:plensJ2}. 
 The second subset consists of the two fiducial SIDM merger models 
  Bw-$\{4,4,12\}$-Bm\_b and By-$\{6,4,12\}$-Bm\_b, for which 
  the final primary's mass is   $\mnw  \sim 10^{15} \msun $.
  The   corresponding X-ray maps and lensing profiles for this 
  second subset are illustrated in Figures \ref{fig:planeJ3} and 
  \ref{fig:plensJ2}, respectively. 
\linebreak

In particular, we highlight  the following points:

(i)  For all the fiducial SIDM merger models,  we find  that 
the observational properties of  El Gordo are  best 
matched  by an  elastic cross-section in the  range  
$\sigma_{11}/{m_X}  \sim 4 \textendash 6 \sxu$. Meanwhile, the inelastic 
up-scattering channel is characterized  by an inelastic cross-section  of 
approximately 
 $\sigma_{12}/{m_X}  \sim 2 \textendash 4 \sxu$ and  a 
threshold velocity of $V_{th} \sim 1200 -1600 \kms$. 
Specifically, we demonstrate that the  presence of an endothermic
reaction channel  is pivotal for  achieving  agreement 
 between the simulated and measured lensing  profiles.
 The latter observational constraint is simultaneously crucial for  ruling 
 out elastic SIDM simulations  as viable merger models.

(ii) The observed twin-tailed X-ray morphology is well reproduced in all  the 
merger models, with the third contour level of the simulated maps  in 
strict agreement with the chosen reference level.  This was achieved  because 
for these SIDM merger models
 we adopted cosmological values for the  initial gas fractions ($ f_g \sim 0.16$).
However, this choice implies   final X-ray luminosities that are higher 
by a factor of $\sim2 $ ($L_X \sim 4\cdot 10^{45} \ergs$)  than the estimated 
value from \citet{Men12}. 
We will return to this issue later.

The SIDM merger models presented here are not without their caveats; 
however, before discussing these limitations, it is worth recalling the primary
motivations for why a 
collisionless CDM merger model remains  difficult to reconcile with the 
observations of the El Gordo cluster. 

(i) As previously outlined in the Introduction, the El Gordo cluster exhibits, 
with high statistical significance, 
an SE  X-ray emission peak that is farther away  from the center of mass than 
the SE DM centroid. This is in contrast  with dissipative arguments and 
differs significantly  from the behavior expected within a collisionless CDM 
framework.
As a solution to this observational feature, \citet{Ng15} introduced a 
returning scenario, in which  the merger is in  a post-apocenter phase and the
two DM halos are 
  moving toward each other, while   the SE X-ray peak  moves in the 
opposite direction.
However, previous findings \citepalias{Valda24}, as well as results of Section 
\ref{sec:opt}, demonstrate that the post-collision X-ray structures  are very 
short-lived ($ \tau_X \sim 0.1\textendash 0.3 $ Gyr)   compared to  orbital  time scales
 ($ \sim 2 $ Gyr).  It is therefore difficult for  such a scenario to reproduce 
  the observed twin-tailed X-ray morphology of the  El Gordo cluster.

(ii)  Finally, it is worth recalling that in a standard CDM merger  
 the  relative line-of-sight peculiar velocity 
between the SE and NW  BCG components is expected  to provide an unbiased
estimator of DM particle velocities.  As shown  in the maps of Figure 
\ref{fig:planeA1},  this
velocity is expected to lie in the range 
$V^s_r\sim 1,200 \textendash 1,500 \kms $, which  
is significantly higher than the measured value of 
$V_r^s\simeq 600 \kms$ \citep{Men12}.

These points clearly support the validity of an SIDM-based  merger model
 for the El Gordo cluster. Nonetheless, there are several critical issues 
 regarding our fiducial SIDM merger models  that warrant  further discussion.

(i) As previously outlined, the final X-ray luminosity in all of our merger 
models is about $L_X \sim 4\cdot 10^{45} \ergs$, which is nearly twice the 
estimated value 
 \citep{Men12}. While  $L_X$ could be reduced  by considering a  lower 
 cluster gas fraction ($f_g \rightarrow 0.1$),  such a reduction   would lead 
 to a lower  overall emission, resulting  in  a  third contour level located 
 much closer to
 the position of the SE X-ray peak.
 Another solution would be to increase  the initial core radius of the 
 SE cool core region, thereby reducing the cuspiness of the central gas density
 peak and, in turn,  the final X-ray luminosity $L_X$. 
 
 Unfortunately, such an approach 
 has a significant drawback: for SIDM merger models with these initial 
conditions,  the final  location of the X-ray peak   now  trails the DM
centroid,   exhibiting a negative  offset $\dxse$.
According to previous findings  \citep{V25}, 
this  occurs because the SE cool core \textendash~ now characterized by a 
larger core 
radius  \textendash~ is  subject to  a larger ram pressure force during the collision, 
leading to a more pronounced deceleration
(see also Section 3.4 of \citetalias{Valda24}).

Another critical issue regarding the proposed two-state, inelastic SIDM merger 
models 
is that the quality of their agreement with the adopted set of El Gordo 
observational 
constraints depends critically on the significance level of the WL error 
measurements.
More specifically, we outline below  the most important points  where 
a future WL analysis of El Gordo \textendash  leading to a significant 
improvement in the statistical errors \textendash will have a crucial impact.

(ii) A critical issue is the angular dependency of the measured lensing profiles
at small angles ($\theta \simlt 50^{''}$).
If future WL measurements   reduce the statistical uncertainty in the 
innermost bin and  confirm an NFW-like behavior for the 
lensing profiles $g_T(\theta)$, this  will pose a significant challenge 
to  the proposed SIDM merger models.
It is worth noting that such a lensing profile would  be 
 problematic  for any SIDM scenario attempting  to explain the observed properties 
of El Gordo: it would imply a physical contradiction wherein the  DM  exhibits 
collisional properties during the cluster-scale encounter, but it remains effectively 
collisionless with respect  to the internal dynamics of the individual  halos.

(iii) Another important aspect  is the statistical significance of the positional errors 
for the WL mass centroids. According to  the WL analysis of \citet{Kim21}, 
 a displacement of $\sim 10^{''} (\sim 80 \kpc)$  corresponds approximately to 
the 2 $\sigma$ uncertainty range  for  the mass centroid positions.
Clearly, if these uncertainties were  reduced by improved  WL measurements, 
this would, in turn, imply   a decrease in  the statistical errors
associated with the various measured offsets.

Specifically, if future WL measurements  confirm  a mass of the primary of 
 $\mnw  \sim  10^{15} \msun $, along with a projected separation 
between the two mass centroids  exceeding
$ {\it d^{obs}_{DM}} \gtrsim 750 \kpc$ and an offset  
$d^{SE}_{\XDM} \gtrsim 50 \kpc $, 
then the two fiducial SIDM merger models of  Figure \ref{fig:planeJ3}
would face  serious difficulties.

For these merger models,  Bw-$\{4,4,12\}$-Bm\_b and By-$\{6,4,12\}$-Bm\_b,
the simulated X-ray maps in the top panels of Figure \ref{fig:planeJ3} indicate that the 
best match to the data is achieved when $d_{DM} \simlt 710 \kpc $ 
 and $d^{SE}_{\XDM} \simlt 40 \kpc $. As  previously discussed  in 
Section \ref{sec:opt}, an increase in   $d_{DM}$ 
implies a corresponding increase  in both  $d^{SE}_{\XDM} $ and $\tpcl$; 
however, this increase would also introduce a one-tailed X-ray morphology 
 at the present epoch, which is in sharp contrast to  what is observed.
 Finally, it is worth noting that a statistical unambiguous detection of 
  a positive BCG-DM offset exceeding $d_{\BDM} \simgt 50 \kpc $   would 
 clearly put all the two-state SIDM merger models 
 presented here  in jeopardy.

  As outlined in the Introduction, the study presented here   
  considers a two-state, endothermic SIDM model. 
 It must be stressed that our findings are strictly dependent on 
 the adoption of this specific framework. Allowing for a more general 
 SIDM model would imply the presence of additional  scattering channels. 
 These could include  \citep{ON23} either an elastic scattering channel
 between the ground state and the excited  state 
 ($\chi_1+\chi_2  \rightarrow  \chi_1+\chi_2$)  or an inelastic 
 down scattering channel ($\chi_2+\chi_2  \rightarrow  \chi_1+\chi_1$).

 In both cases the presence  of these scattering channels,  
 with interaction  cross sections in the same range 
 ($ \sim 2 \textendash 6 \sxu$) as those  considered here, is expected  to 
 significantly modify the inner density profiles of the post-collision DM halos.
These reactions would lead   to cored DM density 
 profiles in the inner regions of the halos \citep{Vog19}, thereby  producing 
  decreasing shear lensing profiles  at small angles.
 This tendency toward smaller shear values at small angles would diverge from
 the  observed behavior (see  also Figure 3 of \citet{V25}).

 Similar considerations apply to SIDM models with  
 velocity-dependent cross-sections. Such models were originally proposed 
 by \citet{Kap16}, who postulated a DM cross-section $\sigma_{DM}(v)/m_X$ that
 falls with relative velocity $v$ to account for the diversity of 
 density profiles from galaxy-size objects up to cluster scales.

 For these SIDM models, the current upper limit of 
 $\sigma_{DM}/{m_X}  \simlt 1 \sxu$ on cluster scales \citep{Rob19} is
 clearly in conflict with our findings,  which indicate that the measured 
 cluster offsets can hardly be reproduced by simulations satisfying these 
 constraints (see below). Notably, increasing the scattering cross-section
 to the required level (${\sigma_{DM}}/{m_X} \sim 4 \textendash 6\sxu$) 
 would  imply raising the DM cross-section at dwarf-sized scales 
 to uncomfortably high levels (e.g.  $\sigma_{DM}/m_X\sim 10 \sxu$).

 Finally, a significant limitation of  
 our fiducial SIDM merger models is that the best match  to the chosen set of
 El Gordo observational constraints is obtained  
 for an elastic  DM scattering cross-section in the range 
${\sigma_{11}}/{m_X} \sim 4 \textendash 6\sxu$. Such an interval is clearly
inconsistent with the present  upper limits  
(${\sigma_{DM}}/{m_X} \simlt 1 \sxu$) derived by various authors
 \citep{Rob17,Rob19,Kim17,Wittman18,Harvey19,Shen22,Cross24,Roche24} 
on  galaxy  cluster scales.

The Bullet Cluster represents a critical case study for this tension.
For this merging system, observational upper limits on the 
galaxy-DM offset \citep[$\simlt 20 \kpc$, ][]{Randall08}  have been utilized
by many authors \citep{Mark04,Randall08,Rob17} 
to derive  upper limits on the order of ${\sigma_{DM}}/{m_X} \simlt 1 \sxu$
for the DM cross-section in elastic SIDM merging simulations.
Such values are clearly incompatible with our findings.

The discrepancies arising from these  two different findings were already
discussed in the Conclusions of \citetalias{Valda24}, 
where it was proposed that this conflict could be resolved by  assuming  
that a key parameter is the collisional energy of the merger.
Specifically, it was  argued that,  during a cluster merger,   the  
ability of DM to  exhibit collisional properties  
 is strongly correlated  with  the energy of the cluster collisions. 

 In the proposed scenario, the behavior of DM during a cluster merger
 remains collisionless unless the collisional energy of  the two colliding
 DM halos  exceeds a specific energy threshold, $E_{\rm crit}$.
 As  found in \citetalias{Valda24},   the collisional energy  for the
 El Gordo cluster  is estimated to be 
 around $ E_{\rm{EG}} \sim 1.4 \cdot 10^{64} {\rm{\,erg}} $, 
 while for the  Bullet Cluster, one obtains 
$ E_{\rm{Bullet}} \sim 3 \cdot 10^{63} {\rm{\,erg}} $.

These estimates  suggest $ E_{\rm crit} \sim E_{\rm{EG}} $,  and the proposed 
scenario would then be significantly supported by  measurements of
substantial galaxy-DM offsets (e.g. $\sim 100 \kpc$) in other major merging 
clusters, as massive as El Gordo. To this end,  we  refer to 
 Table 2 of \citet{Kim17},  in which the authors report
 galaxy-DM offsets, as well as  other measurements, for a set 
of known near-equal-mass cluster binary mergers.

According to this table, in addition to El Gordo,  the Sausage Cluster
CIZA J2242.8+5301 at $z=0.19$ appears as the only  promising candidate 
that  can satisfy the required constraints.
The X-ray morphology of this cluster favors a nearly  head-on 
collision, with a collision plane close to the plane of the sky \citep{Daw15}.
For this cluster  the total mass is estimated  to be 
$\sim 2\cdot 10^{15} \msun$, with the two subclusters  
having approximately equal mass and the  corresponding 
DM centroids  separated by about $\sim 1 \mpc$ \citep{Jee15}.

The galaxy-DM offsets are found to lie between $\sim 50$ and 
$\sim 300 \kpc$, although  with large uncertainties. 
It is worth noting that in the northern group, as expected in an SIDM scenario,
 the  galaxy centroid appears farther away  from the system center of mass than 
 the corresponding mass peak. Such an offset is statistically 
 significant \citet{Jee15}, at variance with that of the southern cluster,
for which  the errors are relatively large.  

We can estimate the collisional  energy of the merger 
by adopting the collision parameters  of \citet{Jee15}, yielding 
 $ E_{\rm{Sausage}} \sim 1.5 \cdot 10^{64} {\rm{\,erg}} $.
 A key aspect of this assessment is that we obtain
 $ E_{\rm{Sausage}} \sim E_{\rm{EG}} $, therefore lending support 
 to the suggested hypothesis  that in massive merging clusters 
 the collisional properties of DM 
 come into play for  collisional energies  
 $ E_{\rm crit} \simgt  10^{64} {\rm{\,erg}} $
 \citepalias{Valda24}.

 Finally, it is worth recalling  that  for merging clusters of 
 $\sim 10^{15} \msun$,   the galaxy-DM offset predicted  by  SIDM merging 
 simulations  with ${\sigma_{DM}}/{m_X} =1  \sxu$  is $\simlt 20 \kpc$ 
 \citep{Kim17}; which is approximately 
 an order of magnitude smaller than the measured offset for the Sausage Cluster.

A viable theoretical framework  that can provide a physically motivated 
description of  such DM behavior \textendash~exhibiting significant collisional
properties in high-energy cluster collisions while remaining  nearly
collisionless at lower energies \textendash~is that of resonant 
SIDM (rSIDM).

Over the years,  there has been a growing interest  
in this class of SIDM models
\citep{Chu19,Csa22,Ben23,Gil23,Kam24,Bel25,Che25,Tra25}. 
For an rSIDM model to successfully simulate the El Gordo merging cluster, 
it must possess a DM cross-section $\sigma_{DM}(E)$ 
with a resonance 
strongly peaked around $ v\sim 3,500 - 4,000\kms$.

This range of  values for $v$ corresponds  approximately to the relative 
velocity between  the two subclusters 
at the pericenter, for the range of collision parameters 
\{ $\mnw,~\mse,~ V, ~P\}$ considered here.
Note that such an energy dependence for  $\sigma_{DM}(E)$  
is fundamentally different from that of previously 
proposed rSIDM models (see, e.g., Figure 1 of \citet{Chu19}).
This class of models appears very promising for reconciling the discussed 
tension between the upper bounds on the DM cross-section at different cluster
scales;  we therefore postpone the study of the El Gordo cluster in an 
rSIDM scenario to  future work.

Finally, based on  the findings presented here, we conclude 
that SIDM merger models more general than purely elastic ones can better 
 characterize the  behavior of DM during cluster  collisions.
  In particular, we argue that future lensing-based observations of massive, 
  high-velocity merging clusters  may  hold the key to uncovering  the true
  nature of DM.

%--------------------------------------------------
%
% PRESS RELEASE : Ramani
\section*{Acknowledgements}
 The simulations presented in this paper were performed   using the  
the Galileo cluster at the CINECA Supercomputing Centre in Bologna, Italy.
CPU time was provided through an SISSA-CINECA agreement and  an ISCRA Category B grant.
The author is also grateful to the anonymous referee for
constructive comments  that  improved the presentation of the paper.

\section*{Data Availability}
The data underlying this article will be shared on reasonable request
to the corresponding author.

%--------------------------------------------------
%----------------- BIBLIOGRAPHY -------------------


\begin{thebibliography}{}


\bibitem[Alvarez  \&  Yu (2020)]{Alv20}
Alvarez, G. \&  Yu, H.B.,  \ 2020, \prd, 101, 043002


\bibitem[Asencio et al. (2021)]{As21}
{Asencio}, E.,   {Banik}, I.  \& {Kroupa}, P.,  \ 2021, \mnras, 500,  5249

\bibitem[Asencio et al. (2023)]{As23}
{Asencio}, E.,   {Banik}, I.  \& {Kroupa}, P.,  \ 2023, \apj, 954 , 162


\bibitem[B{\'e}langer et al. (2025)]{Bel25}
B{\'e}langer, G., Chakraborti, S.,  Delaunay, C., et al., \ 2025,  \prd, 112,
 095039

\bibitem[Beneke et al. (2023)]{Ben23}
Beneke, M., Lederer, S. \& Urban, K. \ 2023, Phys. Lett. B, 839, 13777

\bibitem[Burkert (1995)]{Bu95}
Burkert, A., \ 1995, \apj, 447, L25 


\bibitem[Chadayammuri et al. (2022)]{Cha22}
{Chadayammuri}, U., ZuHone, J., Nulsen, P.  et al., \ 2022, \mnras,  {509}, 1201

\bibitem[Cheng et al. (2025)]{Che25}
Cheng, Y., Ge, S.-F., Sheng, J. et al., \ 2025,  Phys. lett. B, 861, 139290

\bibitem[Chu et al. (2019)]{Chu19}
Chu, X., Garcia-Cely, C. \&  Murayama, H.,  \ 2019, Phys. Rev. Lett., 122, 071103


\bibitem[Chua et al. (2021)]{Chua21}
Chua, K.T.E., Dibert, K., Vogelsberger, M.,  et al.,  \ 2021, \mnras, 500,  1531


\bibitem[Cross et al. (2024)]{Cross24}
Cross, D., Thoron, G., Jeltema, T.~E., et al., \ 2024, \mnras, 529, 52

\bibitem[Cs{\'a}ki et al. (2022)]{Csa22}
Cs{\'a}ki, C., Gomes, A., Hochberg, Y.  et al., \ 2022, 
 Journal of High Energy Physics, 11, 162


\bibitem[Dawson et al. (2015)]{Daw15}
Dawson, W.~A.,  {Jee}, M.~J.,  {Stroe}, A. et al., \ 2015, \apj, 805 143

\bibitem[Diego et al. (2020)]{Die20}
Diego, J.~M.,  {Molnar}, S.~M.,  Cerny, C.,  et al. \ 2020, \apj, 904, 106

\bibitem[Diego et al. (2023)]{Die23}
Diego, J.M.,  Meena, A.K.,  Adams, N.J., et al., \ 2023,  \aap, 672, A3


\bibitem[Donnert (2014)]{Donnert14}
Donnert, J.~M.~F.,  \ 2014, \mnras, 438, 1971

\bibitem[Donnert et al. (2017)]{Donnert17}
Donnert, J.~M.~F., {Beck}, A.~M., Dolag, K., et al. \ 2017, \mnras, 471, 4587


\bibitem[Drakos et al. (2017)]{Drakos17}
Drakos, N.~E., Taylor, J.~E. \& Benson, A.~J., \ 2017, \mnras, 467, 2345


\bibitem[Duffy et al. (2008)]{Du08}
Duffy, A.~R., Schaye, J., Kay, S.~T., et al.  \ 2008, \mnras, 390, L64


\bibitem[Fischer et al. (2022)]{Fis22}
Fischer, M.~S., Br{\"u}ggen, M., Schmidt-Hoberg, K. et al., \ 2022,  \mnras, 510, 4080

\bibitem[Fischer et al. (2023)]{Fis23}
Fischer, M.S., Durke, N.H., Hollingshausen, K., et al. \ 2023, \mnras, 523, 5915

\bibitem[Gilman et al. (2023)]{Gil23}
Gilman, D., Zhong, Y.-M. \&   Bovy, J., \ 2023, \prd, 107, 103008


\bibitem[Gregory \& Thompson (1984)]{Greg84}
Gregory, S.~A. \& Thompson, L.~A., \ 1984, \apj, 286, 422


\bibitem[Harvey et al. (2014)]{Harvey14}
Harvey, D., Titley, E.,  Massey, R. et al., \ 2014, \mnras, 441, 404


\bibitem[Harvey et al. (2019)]{Harvey19}
Harvey, D., Robertson, A.,  Massey, R., et al.  \ 2019, \mnras, 488, 1572


\bibitem[Huo et al. (2020)]{Huo20}
Huo, R., Yu, H.-B. \& Zhong, Y.-M., \ 2020, J. Cosmol. Astropart. Phys., 06, 051



\bibitem[Jee et al. (2014)]{Jee14}
Jee, M.~J., {Hughes}, J.~P., {Menanteau}, F., et al. \ 2014, \apj , 785, 20


\bibitem[Jee et al. (2015)]{Jee15}
 {Jee}, M. J.  Stroe, A., Dawson, W., et al.  \ 2015, \apj, 802,  46


\bibitem[Kahlhoefer et al. (2014)]{Kah14}
Kahlhoefer, F.,  Schmidt-Hoberg, K., Frandsen, M.~T., et al.  \ 2014,
\mnras, 437, 2865


\bibitem[Kamada \& Kim (2024)]{Kam24}
Kamada, A. \&  Kim, H.-J.,  \ 2024, \prd, 109, 063535


\bibitem[Kim et al. (2017)]{Kim17}
Kim, S.~Y., Peter, A.~H.~G. \& Wittman, D. et al., \ 2017,  \mnras, 469, 1414 

\bibitem[Kim et al. (2021)]{Kim21}
Kim, J., Jee, M.~J.,  Hughes, J.~P.  et al. , \ 2021, \apj, 923, 101


\bibitem[Kim et al. (2025)]{Kim25}
Kim, J.H., Kong, K.,  Lim, S.H., et al. \ 2025, J. Cosmol.  Astropart.  Phys.,  04, 016,

\bibitem[Kaplinghat et al. (2016)]{Kap16}
Kaplinghat, M.,  Tulin, S.  \&  Yu, H.-B., \ 2016, Phys. Rev. Lett., 116, 041302

\bibitem[Kravtsov et al. (2018)]{Kr18}
Kravtsov, A.~V., Vikhlinin, A.~A. \&{Meshcheryakov}, A.~V., \ 2018, Astronomy Letters, 
44, 8


\bibitem[Lage \& Farrar (2014)]{La14}
{Lage}, C. \& {Farrar}, G.~R.,  \ 2014,  \apj,  787, 144



\bibitem[Leonard et al. ( 2024)]{Leo24}
Leonard, A., O'Neil, S., Shen, X.,  et al. \ 2024, \mnras, 531, 1440



\bibitem[Machado \& Lima Neto (2013)]{Mac13}
{Machado}, R.~E.~G. \& Lima Neto, G.~B.,  \ 2013, {\mnras}, 430, 3249


\bibitem[Markevitch et al. (2004)]{Mark04}
Markevitch, M.,  {Gonzalez}, A.~H.,  {Clowe}, D.  et al., \ 2004,
 {\apj}, 606, 819



 \bibitem[Mastropietro \& Burkert  (2008)]{Mas08}
{Mastropietro}, C. \&  {Burkert}, A., \ 2008, {\mnras},  {389}, 967


\bibitem[Merritt et al. (2006)]{Mer06}
{Merritt}, D.,  {Graham}, A.~ W.,  {Moore}, B., et al. \ 2006,  \aj, 132, 2685


\bibitem[Molnar \& Broadhurst  (2015)]{Molnar15}
Molnar, S.~M. \& {Broadhurst}, T., \ 2015, \apj, 800, 37

\bibitem[Molnar \& Broadhurst  (2017)]{Mol17}
Molnar, S.~M. \& {Broadhurst}, T., \ 2017, \apj, 841, 46


\bibitem[Moura et al. (2021)]{Mou21}
Moura, M.~T.,  Machado, R.~E.~G. \& {Monteiro-Oliveira}, R., \ 2021, \mnras, 500, 1858


\bibitem[Menanteau et al. (2012)]{Men12}
{Menanteau}, F., Hughes, J.~P., {Sif{\'o}n}, C.,  et al., \ 2012, \apj, 748, 7


\bibitem[Molnar(2016)]{Molnar16}
 {Molnar}, S.,  \ 2016 , Front. Astr. Space Sci.,  2, 7 


\bibitem[Ng et al. (2015)]{Ng15}
{Ng}, K.~Y.,  {Dawson}, W.~A., {Wittman}, D.,  et al., \ 2015, \mnras, 453, 1531

\bibitem[O'Neil et al. (2023)]{ON23}
O'Neil, S., Vogelsberger, M., Heeba, S., et al. \ 2023, \mnras, 524, 288

\bibitem[Poole et al. (2006)]{Poole06}
{Poole}, G.~B., {Fardal}, M.~A.,  {Babul}, A., et al., \ 2006, {\mnras}, 373, 88

\bibitem[Power et al. (2003)]{Pow03}
Power, C.,  Navarro, J.~F.,  Jenkins, A.  et al, \ 2003, \mnras, 338, 14


\bibitem[Price (2012)]{Price2012}
Price, D. J.,  \ 2012, J. Comp. Phys., 231, 759

\bibitem[Randall et al. (2008)]{Randall08}
Randall, S.~W., Markevitch, M,  Clowe, D., Gonzalez, A.~ H. \&
Brada{\v{c}}, M., \ 2008, \apj, 679, 1173


\bibitem[Ricker (1998)]{Ricker98}
Ricker, P.~M.,  \ 1998, \apj, 496, 670

\bibitem[Ricker \& Sarazin (2001)]{Ricker01}
{Ricker}, P.~M. \&  {Sarazin}, C.~ L.,  \ 2001, \apj, 561, 621


\bibitem[Robertson et al. (2017)]{Rob17}
Robertson, A.,  Massey, R. \& Eke, V.,  \ 2017, \mnras, 465, 569


\bibitem[Robertson et al. (2019)]{Rob19}
Robertson, A., Harvey, D., Massey, R.  et al., \ 2019,  \mnras, 488, 3646


\bibitem[Roche et al. (2024)]{Roche24}
Roche, C., McDonald, M.,  Borrow, J., et al. \ 2024, Open Journal of 
Astrophysics, 7, 65


\bibitem[Sabarish et al. (2024)]{Sab24}
Sabarish, V.M., Br{\"u}ggen, M.,  Schmidt-Hoberg, K., et al. \ 2024, \mnras, 
529, 2032

\bibitem[Schutz  \& Slatyer  (2015)]{Sch15}
Schutz, K. \&  Slatyer, T.R., \ 2021, J. Cosmol. Astropart. Phys., 01, 021

\bibitem[Secco et al. (2018)]{Secco18}
Secco, L.~F.,  Farah, A., Jain, B., et al.,  \ 2018, \apj, 860,  32


\bibitem[Sheardown et al. (2019)]{Sh19}
Sheardown, A., Fish, T.~M., Roediger, E., et al. \ 2019, \apj, 874, 112


\bibitem[Shen et al. (2022)]{Shen22}
Shen, X., Brinckmann, T.,  Rapetti, D.,  et al. \ 2022, \mnras, 516, 1302

\bibitem[Shen et al. (2024)]{Shen24}
{Shen}, X., {Hopkins}, P.F.,  {Necib}, L., et al.  \ 2024, \apj, 966, 131


\bibitem[Sirks et al. (2024)]{Sirks24}
Sirks, E.L., Harvey, D., Massey, R., et al.  \ 2024, \mnras, 530, 3160


\bibitem[Springel \& Farrar (2007)]{Spr07}
Springel, V.  \& {Farrar}, G.~R.,  \ 2007,  \mnras,  380, 911

\bibitem[Todoroki \& Medvedev (2019)]{Todo19}
Todoroki,K. \&    Medvedev, M.V., \ 2019, \mnras, 483, 4004

\bibitem[Todoroki \& Medvedev (2022)]{Todo22}
Todoroki,K. \&    Medvedev, M.V., \ 2022, \mnras, 510, 4249

\bibitem[Tran et al. (2025)]{Tra25}
Tran, V., Shen, X.,  Gilman, D. et  al., \ 2025,  \prd, 112, 083003


\bibitem[Tulin \& Yu (2018)]{Tulin18}
{Tulin}, S. \& {Yu}, H.-B.,   \ 2018, \physrep, 730,  1


\bibitem[Umetsu (2020)]{Um20}
Umetsu, K., \ 2020,  Astr. \& Astrophys. Review, 28, 7 


\bibitem[Valdarnini (2016)]{V16}
Valdarnini, R., \ 2016, ApJ, 831, 103


\bibitem[Valdarnini \& Sarazin (2021)]{VS21}
Valdarnini, R. \&  {Sarazin}, C.~L., \ 2021, {\mnras}, {504}, 5409

\bibitem[Valdarnini (2024)]{Valda24}
	Valdarnini, R., \ 2024,  Astr. \& Astrophys., 684, A102, 
	\citepalias{Valda24}

\bibitem[Valdarnini (2025)]{V25}
Valdarnini, R., \ 2025,  arXiV:2509.01298

\bibitem[Vogelsberger  et al. (2012)]{Vog12}
Vogelsberger, M., Zavala, J. \&  Loeb, A., \ 2012, \mnras, 423, 3740

\bibitem[Vogelsberger  et al. (2019)]{Vog19}
{Vogelsberger}, M.,  Zavala, J., Schutz, K., et al.  \ 2019, \mnras, 484, 5437


\bibitem[Weinberg et al. (2015)]{Wei15}
Weinberg, D.H., Bullock, J.S., Governato, F., et al. \ 2015, 
Proc. Nat. Acad. Sci., 112, 12249


\bibitem[Wittman et al. (2018)]{Wittman18}
Wittman, D., Golovich, N.  \&  Dawson, W.~A., \ 2018, \apj, 869, 104


 \bibitem[Zhang et al. (2015)]{Zh15}
 {Zhang}, C.,  {Yu}, Q. \&  {Lu}, Y.,  \ 2015, \apj, 813,  129 

 \bibitem[Zhang et al. (2018)]{Zh18}
{Zhang}, C.,  {Yu}, Q. \&  {Lu}, Y.,  \ 2018, \apj, 855,  36

\bibitem[Zitrin et al. (2013)]{Zi13}
{Zitrin}, A., {Menanteau}, F., {Hughes}, J.~P., et al. \ 2013, \apjl, 770, L15



\bibitem[ZuHone  et al.  (2019)]{ZuH19}
{ZuHone}, J.~A., Zavala, J. \& Vogelsberger, M., \ 2019, \apj, 882, 119

\end{thebibliography}
\end{document}